\documentclass{elsarticle}
\usepackage[margin=2.5cm]{geometry}
\usepackage{setspace}
\usepackage{amssymb}
\usepackage{extarrows}
\usepackage{hyperref}
\hypersetup{colorlinks=true,allcolors=ForestGreen}
\usepackage{tabularx}
\usepackage{booktabs}
\usepackage{literate}
\usepackage{subcaption}
\usepackage{yfonts}
\usepackage{fancybox}
\usepackage{enumitem}
\usepackage{envelope}
\usetikzlibrary{external}
\pgfplotsset{every axis legend/.append style={draw=none}}
\lstMakeShortInline[columns=fixed]!
\newcommand{\arate}[2][{1cm}]{\parbox{#1}{\centering\scriptsize $#2$}}

\begin{document}
\title{
Rule-based epidemic models
}
\author[1,2]{W.~Waites\fnref{cso,stam}\corref{cor}}
\ead{wwaites@inf.ed.ac.uk}
\author[3]{M.~Cavaliere}
\author[4]{D.~Manheim}
\author[5,6]{J.~Panovska-Griffiths\fnref{nihr}}
\author[1,7]{V.~Danos}

\address[1]{\small School of Informatics, University of Edinburgh, Edinburgh, UK}
\address[2]{\small Centre for Mathematical Modelling of Infectious Diseases,
  London School of Hygiene and Tropical Medicine, London, UK}  
\address[3]{\small Department of Computing and Mathematics, Manchester Metropolitan University, Manchester, UK}
\address[4]{\small University of Haifa Health and Risk Communication Research Center, Haifa, Israel}
\address[5]{\small Department of Applied Health Research, University College London, London, UK}
\address[6]{\small Institute for Global Health, University College London, London, UK}
\address[7]{\small The Queen's College, University of Oxford, Oxford, UK}
\address[8]{\small D\'{e}partement d'Informatique, \'{E}cole Normale Sup\'erieure, Paris, France}
\cortext[cor]{Corresponding author}
\fntext[cso]{Supported by the Chief Scientist Office grant number COV/EDI/20/12}
\fntext[stam]{Supported by the Medical Research Council (MRC) grant number MR/V027956/1}
\fntext[nihr]{Supported by the National Institute for Health Research (NIHR)
  Applied Health Research and Care North Thames at Bart’s Health NHS Trust (NIHR
  ARC North Thames)}


\begin{abstract}
Rule-based models generalise reaction-based models with reagents that have internal state and may be bound together to form complexes, as in chemistry. An important class of system that would be be intractable if expressed as reactions or ordinary differential equations can be efficiently simulated when expressed as rules. In this paper we demonstrate the utility of the rule-based approach for epidemiological modelling presenting a suite of seven models illustrating the spread of infectious disease under different scenarios: wearing masks, infection via fomites and prevention by hand-washing, the concept of vector-borne diseases, testing and contact tracing interventions, disease propagation within motif-structured populations with shared environments such as schools, and superspreading events.  Rule-based models allow to combine transparent modelling approach with scalability and compositionality and therefore can facilitate the study of aspects of infectious disease propagation in a richer context than would otherwise be feasible.
  

\end{abstract}

\maketitle



\section{Introduction}

Compartmental models in epidemiology are mathematically equivalent to, and can be expressed in the same way as, chemical reaction models.
The classical model of Kermack and McKendry~\cite{kermack_contribution_1927}, for example, can be written,
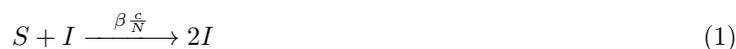
\begin{align}
  \label{eq:kkinfection}
  S + I &\xlongrightarrow{\arate{\beta\frac{c}{N}}} 2I\\
  I &\xlongrightarrow{\gamma} R
\end{align}
This represents infection as an interaction between a susceptible individual and an
infectious one that results in two infectious individuals at rate $\frac{\beta
  c}{N}$ and the recovery or removal of an infectious individual at rate
$\gamma$.
Kermack and McKendry derive differential equations for the case where the rates are constant from first principles and arrive at a system describing the changing quantities of individuals of each kind.
These differential equations are identical to those obtained by considering the above as a chemical reaction system, interpreting $S$, $I$, and $R$ as chemical species in place of exclusive subpopulations -- it is perhaps not a coincidence that Kermack was a biochemist.

This class of model is still in current use~\cite{anderson_infectious_1992,diedre2009,heesterbeek2015}.
To represent the natural history of a particular disease, compartments may be added.
It is common to add a latent compartment for those individuals that are infected but not yet infectious.
Sometimes several kinds of infectious compartments are used to represent different severities or stages of disease progression.
However, in typical differential equation modelling, increasing the number of compartments comes at the cost of poor
scaling: the number of possible interactions increases with the square of the
compartments.
It is just possible to accommodate the compartment explosion for age-stratified
models~\cite{rohani_contact_2010}.
Dividing the population into 8 age bands requires 64 interactions to capture
infection and 8 more for removal.
All of the other transitions for disease progression, from latent to infectious,
among the various severities of infectiousness, result in a modest increase of 8
each.

Explicitly enumerating stratified compartments begins to become unwieldy when other features that would arbitrarily subdivide the population further, or when multiple geographic regions~\cite{eubank_modelling_2004} are considered.
Isolation, removal from, or attenuation of participation in the infection dynamics as a result of testing, clinical diagnosis, or simple precaution, induces a doubling of compartments: we must account for the possibility of both isolated and unconfined individuals of each kind. 
The wearing or not of masks has a similar consequence, as does introducing any feature that subdivides the population, or includes multiple populations.
For example, when we consider the simplest Susceptible-Infected-Recovered (SIR) model, it scales as shown in Table~\ref{tab:features}.
\begin{table}[ht]
\centering
    \begin{tabularx}{0.4\textwidth}{@{} l | rrrr @{}}
      features     & 1 & 2  & 3  & 4\\
      compartments & 6 & 12 & 24 & 48\\
      transitions  & 6 & 20 & 72 & 272\\
    \end{tabularx}
    \par\vspace{0.5\baselineskip}
    \caption{The increase in compartments and transitions with the addition of
      features.}
    \label{tab:features}
\end{table}

The reason for this large increase in the number of compartments and required
transition rates is easily seen.
This formulation requires compartments to represent disjoint subsets of the
population, and this, in-turn, implies redundant specification of interactions
where they are independent of the features.
The progression from latent to infectious, for example, is independent of
whether or not one is wearing a mask but one nevertheless must specify these
cases separately; interactions with peers at school have little to do with the structure and composition of one's family (at least to a first approximation).

This phenomenon also has a negative effect on the ability to inspect and understand reaction-based models.
Even if a large model with dozens of compartments and hundreds of reactions is correct, and even if it is available for inspection, there is little hope of understanding the reasoning behind the model.
There is also little hope of verifying that the model as written in code is the same as the model that is written in the paper about the model (or, rather, that both are representations of the same abstract model).

There are several strategies used in epidemiological modelling to make some progress in the face of the scaling difficulties posed by adding features to compartmental models~\cite{walters_modelling_2018}. 
A simple approach is to assert that the additional features that do not alter the
structure of the model,  e.g. that wearing a mask, for example, reduces the infection rate~\cite{brienen_effect_2010,tracht_mathematical_2010} or that contact tracing causes infectious individuals to become isolated at some rate~\cite{gumel_modelling_2004,giordano_modelling_2020}. 
Doing this is not to study contact tracing or masks and the interactions of individuals wearing
them, or not, but to presuppose that the effect of these interactions is uniform
and can be captured in a single scalar parameter.
A more sophisticated approach is to forego the elegance of the chemical reaction
or compartmental formulation entirely and explicitly model the individuals in
the population as agents interacting arbitrarily as in individual- or agent-based models~\cite{keeling_individual-based_2000,patlolla_agent-based_2006,marshall_formalizing_2015,hunter_taxonomy_2017,willem_lessons_2017,tracy_agent_2018}. 
This has the opposite problem: where reducing interactions to a scalar is
oversimplification, allowing completely arbitrary interactions brings with it
little analytical or structural insight. 
Agent-based models also have the drawback of needing to specify a large number of assumptions.
The quantity of assumptions often implies too many parameters to be reasonably informed from data or fitting.
The shortcomings of both of these strategies are a result of the choice of
level of abstraction: one too coarse, and the other too fine.

In this paper, we present an alternative approach and show that \emph{rule-based
 modelling}~\cite{danos_formal_2004}, already used in modelling molecular biology~\cite{danos_rule-based_2007,chylek_rule-based_2014,kohler_rule-based_2014,bustos_rule-based_2018},  can be used to express scalable and compositional models in a wide range of relevant epidemiological scenarios.

The advantage of rule-based modelling is that allows for explicit representation of entities in a model and their interactions while disregarding features that are
not relevant. The formalism is also parsimonious: minimal extraneous detail is required to
specify the model in machine-readable form for simulation. 
The approach is also transparent: the machine-readable form corresponds closely to the mathematical form resulting in minimal barriers to inspection and verification of models. 

As we demonstrate in the paper, this allows to compose, run and verify computational models and obtain insights for relevant epidemiological scenarios such as the effects of mask wearing in the transmission of respiratory illnesses, passive transmission
by fomites on surfaces or by active vectors such as mosquitoes, testing, tracing
and isolation, as well as populations with a hybrid well-mixed and network
structure, and superspreading events at gatherings. All of the models described
in this paper are available at
\url{https://git.sr.ht/~wwaites/rule-epi}.

The main contribution of this paper is then to provide a new arrow in the quiver of epidemiological modelling.
Rule-based modelling is expressive enough to capture features of disease transmission and interventions that would be impractical to represent in compartmental models.
At the same time, the language is sufficiently clear to make the individual mechanisms and interactions explicit and subject to examination and review in a way that is rarely feasible even with the best agent- or individual-based models.
We demonstrate the proposed approach by presenting specific models for various phenomena of interest for infectious disease modelling.

\section{Rule-Based Approach}


The \emph{chemical master equation} gives the time-evolution of the distribution of configurations of such a system, the trajectory of distributions~\cite{gillespie_rigorous_1992,anderson_continuous_2011}.
Differential equation formulations such as the one derived by Kermack and McKendry approximate the mean number of each chemical species as a function of time, and this approximation becomes increasingly accurate as this number goes to infinity. There exist methods for obtaining approximate differential equations for the higher moments as well.
\textbf{Rule-based modelling} generalises this by allowing chemical species, rather than being atomic entities, to have internal structure and bonds between particles.

Rule-based formulations, like reaction-based ones have a useful property: compositionality~\cite{blinov_complexity_2008,mallavarapu_programming_2009}.
One can derive differential equations from reactions using the rate equation, a sum over all reactions~\cite{plotkin_calculus_2013,baez_quantum_2018}.
Adding rules is simply adding more terms to this sum (the same is not true for reactions because it is necessary to account for each combination of reagents).
This compositional property of rule-based models means that it is possible to design models in such a way that they can be combined. 
For example, one may combine a model of the flu with one of \textsc{covid}-19, for example, by simply concatenating them.
This is powerful capability has been emphasised in the closely related
Petri net formulation~\cite{baez_open_2020,halter_compositional_2020}.
The advantage of rules over chemical reactions in this connection is ease of variation: a single change to a rule can cascade to many changes in the corresponding reactions~\cite{danos_agile_2009}.

The entities in rule-based modelling are called \emph{agents}.
These agents should not be confused with the agents as they are in agent- or
individual-based modelling in the epidemiology literature; they are much
simpler and they have a precise definition~\cite{danos_formal_2004}.
There are several computer languages for writing rule-based models, the most
well-known are the $\kappa$ language as implemented by the
KaSim~\cite{boutillier_kappa_2020} simulator and the
BioNetGen language~\cite{harris_bionetgen_2016}.
We will use $\kappa$, and introduce the main features of the language here.

An agent with internal states is specified as follows -- in text and
equivalently in the language of KaSim,
\begin{equation*}
  P(x_u), \: u \in \{S, E, I, R\}
\end{equation*}
\begin{center}
  !
\end{center}
\vspace{0.5\baselineskip}
The meaning of this is that there is a set of agents, $P()$ that is partitioned
into disjoint subsets, $P(x_S)$, $P(x_E)$, $P(x_I)$, $P(x_R)$.
This agent, $P$, might refer to a population made up of individuals whose
internal state $x$ corresponds to the compartments of an SEIR model -- a standard extension of SIR with an additional latent or \emph{exposed} compartment whose members are infected but not yet infectious.

It is permitted to have more than one kind of internal state.
For example, one could write,
\begin{equation*}
  P(x_u,m_v), \: u \in \{S, E, I, R\},\; v \in \{Y,N\}
\end{equation*}
\begin{center}
  !
\end{center}
\vspace{0.5\baselineskip}
to represent wearing or not of masks.
One can then refer to those infectious individuals wearing masks, $P(x_I,m_Y)$, all
individuals not wearing masks, $P(m_N) = \bigcup_u P(x_u,m_N)$, or all individuals,
$P() = \bigcup_u\bigcup_v P(x_u,m_v)$.
This is a fundamental difference between rule-based models and compartmental or
reaction models: one can refer to specific subsets of agents as required, and
those internal states that are not relevant can simply not be mentioned.

It is also possible to specify bonds between agents.
This is extensively used in molecular biology to represent polymers, chains of
molecules.
Here, we will make light use of this facility to show how a population can be
given structure.
As with internal states, binding sites have names, and bonds are numbered.
\begin{equation*}
  P(c^u),\; u \in \mathbb{N^+}
\end{equation*}
\begin{center}
  !
\end{center}
\vspace{0.5\baselineskip}
The number is arbitrary and meaningful only within the scope of an expression.
Thus, $P(c^1), P(c^1)$, denoting two bound agents, means the same thing as
$P(c^{42}), P(c^{42})$, provided that 42 is not used elsewhere in the same
expression.
An unbound site is written with a $\cdot$: $P(c^\cdot)$.

A rule, much like a chemical reaction, has a left- and a right-hand side.
A rule for infection, again both in text and in the language of KaSim, is,
\begin{equation*}
  P(x_S),\; P(x_I) \xlongrightarrow{k} P(x_I),\; P(x_I)
\end{equation*}
\begin{center}
  !'infection' P(x{s}), P(x{i}) -> P(x{i}), P(x{i}) @ k!
\end{center}
\vspace{0.5\baselineskip}
This is very similar to the chemical reaction representation in
Equation~\ref{eq:kkinfection}, and the formulation in code corresponds exactly
to the more attractively typeset mathematical version.

A convenient shorthand, with no change of meaning, useful when the only
difference between the left- and right-hand side of a rule is a change of state,
is the \emph{edit notation},
\begin{center}
    !'infection' P(x{s/i}), P(x{i}) @ k!
\end{center}
\vspace{0.5\baselineskip}

The output of such models is defined as a set of \emph{observables}.
An observable is a function of time, and is specified in terms of arithmetic
operations on the cardinalities of the sets of agents.
For example,
\par\vspace{\baselineskip}
\begin{minipage}{0.22\textwidth}
  \centering
  $|P(x_I)|$\\
  $|P(x_E) \cup P(x_I)|$\\
  $|P(m_Y)|/|P()|$
\end{minipage}
\begin{minipage}{0.33\textwidth}
  \centering
  !|P(x{i})|!\\
  !|P(x{e})| + |P(x{i})|!\\
  !|P(m{y})|/|P()|!
\end{minipage}
\begin{minipage}{0.54\textwidth}  
  number of infectious individuals\\
  number of infected individuals\\
  fraction wearing a mask
\end{minipage}
\par\vspace{\baselineskip}

The \texttt{KaSim} stochastic simulator samples trajectories~\cite{gillespie_exact_1977} from models written in this language.
The corresponding differential equations can be generated for integration using GNU Octave, Matlab, Mathematica or Maple with the \texttt{KaDE} program for settings where this is appropriate and stochastic effects can be neglected. 
This means a large class of epidemiologically interesting models typically expressed as ODEs can be written, often more compactly and elegantly, in rule-based form.

As said, with some assumptions, this rule-based form admits composition.
Compositionality is a very convenient property for models: it means that they can be easily combined~\cite{blinov_complexity_2008,mallavarapu_programming_2009}.
Provided that all sites and internal states of agents are consistent, composition of rule-based models is simply concatenation.
This follows from the form of the master equation as a sum over rules.
Care must still be taken that the intended semantics are obtained when composing models.
A duplicate rule will obtain at twice the rate that it otherwise would, and this may or may not be the intent.
Rules applying to partly overlapping subsets of agents present a similar but less obvious difficulty for composition, which can be solved with rule refinement~\cite{danos_rule-based_2008}.
With due care, even without using rule refinement, it is possible to create modular models that can be combined to create more complex ones.

The caveat with producing differential equations is that each complex of agents must be treated as a chemical species.
If agents can become bound together in arbitrarily large complexes, this results in arbitrarily many chemical species and a combinatorial increase in the corresponding number chemical reactions.
To manage this, complexes must be truncated at a certain size if differential equations are to be used.
For this paper, we use the stochastic simulation for all models.

\section{Modelling Epidemics with a Rule-Based Approach}

We describe the rule-based approach and language by presenting six models representing phenomena of
interest in infectious disease modelling and that feature different aspects of compositionality and scalability:
\begin{enumerate}
  \item Mask-wearing including a dynamic process where masks become commonly
    worn and are later abandoned as unnecessary
  \item Fomites, where infection is transmitted through contaminated surfaces
    demonstrating the effectiveness of hand-washing
  \item Vector-borne diseases, with a coupled life-cycle model for the vectors
    and control of an epidemic through elimination of habitat
  \item Testing, as a means of identifying infectious individuals who should be
    isolated, with a finite supply of tests produced by a manufacturing facility
  \item Contact tracing, built upon the previous testing model
  \item Schools, conceived of as two infection processes -- interactions among
    children and interactions among the general population --  coupled through a
    family network.
  \item Gatherings where subsets of the population are more or less likely to
    periodically attend gatherings at which contact frequency is much greater
    than normal.
\end{enumerate}

Each model is described and simulated for reasonable illustrative values of the
relevant parameters. 

\subsection{Masks}
We develop a model for mask wearing that uses the above agent and show how the proposed rule-based language allows extensions with the compositional addition of new features.

The corresponding code is provided in Appendix~\ref{code:masks}.
The progression rules are exactly the same as with a
regular compartmental model,
\begin{align}
  P(x_E) &\xlongrightarrow{\alpha} P(x_I)\\
  P(x_I) &\xlongrightarrow{\gamma} P(x_R)
\end{align}

The infection rules are very much like a stratified compartmental model: we
explicitly specify the four combinations: where the susceptible individual is
wearing a mask, or not, and where the infectious one is, or not.
Let $1 - m_{xy}$ be the effectiveness of mask wearing at preventing infection of
a susceptible individual, during a contact with an infectious individual,
with $x$, $y$ indicating the no-mask/mask status of said individuals.
For example,
\begin{equation}
  \label{eq:maskmatrix}
  m_{xy} = \left[
    \begin{array}{cc}
      1   & 0.8\\
      0.4 & 0.2
    \end{array}
  \right]
\end{equation}
stipulating that: $m_{NN} = 1$, and there is no reduction in infection probability; while 
$m_{YY} = 0.2$ means a very substantial reduction in infection probability if both
parties wear masks.
If only one is wearing a mask, the benefit is relatively small if it is the
susceptible individual and significant if it is the infectious one.
The four rules are then given as,
\begin{equation}
  \label{eq:maskinfection}
  P(x_S,m_u),\; P(x_I,m_v) \xlongrightarrow{m_{uv}\beta\frac{c}{N}} P(x_E,m_u),\; P(x_I,m_v)
\end{equation}
where $\beta$ is the infection probability, $c$ is the number of contacts per
unit time, and $N$ is the total population, as usual.

In addition to the above, which might be sufficient to study the circumstance
where a constant fraction of the population wears masks, we incorporate a simple
mechanism for mask wearing to become popular.
We suppose that the more people wear masks, the more likely it is for
individuals to decide to wear and not to wear masks,
\begin{equation}
  \label{eq:maskon}
  P(m_N),\; P(m_Y) \xlongrightarrow{\frac{\mu}{N}} P(m_Y),\; P(m_Y)\\
\end{equation}
This is a purely crowd-based logic and the result of this positive feedback is like an epidemic of masks.
It eventually results in the entire population wearing masks.
This is clearly unrealistic when the outbreak has run its course.
Therefore, we use a second rule with negative feedback but a different rate constant,  
\begin{equation}
  \label{eq:maskoff}
  P(m_Y),\; P(m_N) \xlongrightarrow{\frac{\nu}{N}} P(m_N),\; P(m_N)
\end{equation}

It is possible to notice that if $\mu > \nu$ and there is at least a small number of individuals wearing masks, then mask usage will
simply grow at a rate of $\mu - \nu$, but if the opposite is true, masks will
fall to zero.
More than a simple logic of following the crowd is needed.
We reason that, in addition to observing the behaviour of others, our agents also have access to information about the outbreak itself, perhaps from watching the nightly news, and spontaneously decide to wear or remove a mask proportionally to the current danger,
\begin{align}
    P(m_N) &\xlongrightarrow{\arate[1.25cm]{\sigma p_I}} P(m_Y)\\
    P(m_Y) &\xlongrightarrow{\arate[1.25cm]{\sigma\left(1-p_I\right)}} P(m_N)
\end{align}
where $p_I = \frac{|P(x_I)|}{N}$, or the chance at the current time of any given individual being infectious.

\emph{Remark:} the above four rules show two ways of having a rate proportional to a subpopulation. The first is to have a bimolecular rule and a constant rate.
The second is to have a unimolecular rule and a variable rate.
To achieve a variable rate a calculation is performed after each event.
The difference is largely a matter of taste as the simulator performs substantially the same computation when incrementally computing propensities~\cite{danos_scalable_2007}.

There remains the question of how to choose the rates $\mu$ and $\nu$ and for present purposes we will select them somewhat arbitrarily with wearing masks significantly faster than removing them.
An alternative formulation not requiring this arbitrary choice involves using memory of the recent past, a technique that we also use for contact tracing in Section~\ref{sec:tti}.

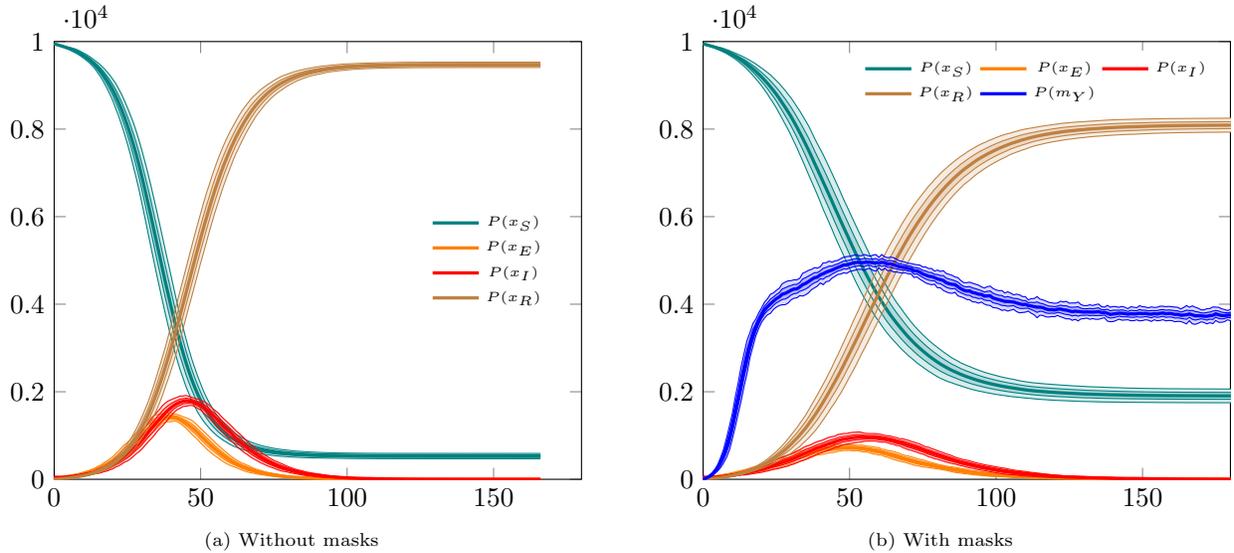
\begin{figure}[ht]
  \begin{subfigure}{0.48\textwidth}
    \resizebox{\textwidth}{!}{
      \begin{tikzpicture}
        \begin{axis}[
          cycle list name=color list, no markers,
          xmin=0, xmax=180, ymin=0, ymax=10000,
          legend style={legend columns=1, at={(0.95,0.5)}, anchor=east},
          ]
          \envelope{teal}{S}{\tiny $P(x_S)$}{data/nomasks.tsv}
          \envelope{orange}{E}{\tiny $P(x_E)$}{data/nomasks.tsv}
          \envelope{red}{I}{\tiny $P(x_I)$}{data/nomasks.tsv}
          \envelope{brown}{R}{\tiny $P(x_R)$}{data/nomasks.tsv}
        \end{axis}
      \end{tikzpicture}
    }
    \caption{Without masks}
  \end{subfigure}
  \hfill
  \begin{subfigure}{0.48\textwidth}
    \resizebox{\textwidth}{!}{
      \begin{tikzpicture}
        \begin{axis}[
          cycle list name=color list, no markers,
          xmin=0, xmax=180, ymin=0, ymax=10000,
          legend style={legend columns=3}, 
          ]
          \envelope{teal}{S}{\tiny $P(x_S)$}{data/masks.tsv}
          \envelope{orange}{E}{\tiny $P(x_E)$}{data/masks.tsv}
          \envelope{red}{I}{\tiny $P(x_I)$}{data/masks.tsv}
          \envelope{brown}{R}{\tiny $P(x_R)$}{data/masks.tsv}
          \envelope{blue}{M}{\tiny $P(m_Y)$}{data/masks.tsv}
        \end{axis}
      \end{tikzpicture}
    }
    \caption{With masks}
  \end{subfigure}
  \caption{%
    The figure shows the results of a SEIR model with masks. Wearing masks
    reduces the probability of transmission by different amounts depending on
    whether the susceptible or infectious individual, or both or neither, are
    wearing masks. Individuals wear masks according to a purely crowd-based
    logic: the more individuals are wearing them, the more likely it is for one
    to make the transition from not wearing to wearing a mask and vice-versa.
    This is supplemented with spontaneous mask wearing and removal proportional to the fraction of the population that is infectious.
  }
  \label{fig:masks}
\end{figure}
The result of running this model as described, and setting $\sigma = 0.5$, are shown
in Figure~\ref{fig:masks}.
With these simple assumptions about the effect of wearing masks, and a direct
implementation of the relevant interactions, we can see that they do have a
significant effect on reducing both the peak number of infections and the
total.
We can also observe that the system settles at an equilibrium of mask wearing.
It is possible to work out precisely the nature of this equilibrium.
Since there are, at equilibrium, very few infectious individuals, there is very little spontaneous mask wearing, and spontaneous removal happens at a rate of $\sigma$.
Masks are also removed due to the crowd logic of Equation~\ref{eq:maskoff}.
At equilibrium these two processes must balance with the crowd logic of Equation~\ref{eq:maskon} causing masks to be worn.

\subsection{Hand washing}

Infection due to contact surfaces contaminated by viral particles that have been
shed (fomites) is said to be mitigated by hand washing.
We model this phenomenon as follows with code in Appendix~\ref{code:fomites}.
Individuals in this model have hands.
Hands can be clean or dirty.
They become clean through washing, and become spontaneously dirty after some
time.
Our agents, therefore, have the signature,
\begin{align}
  P(x_u,h_v),&\; u \in \{S, E, I, R\},\; v \in \{C, D\}\\
  S(c_w),&\; w \in \{Y, N\}
\end{align}
The washing and dirtying of hands are described by the rules,
\begin{align}
  P(h_D) &\xlongrightarrow{\omega} P(h_C)\\
  P(h_C) &\xlongrightarrow{\tau} P(h_D)
\end{align}
where $\omega$ is the rate of hand washing, and $\tau$ is the rate at which
hands become dirty.

Contamination of surfaces is straightforward and the logic is very similar to 
infection in the standard model,
\begin{equation}
  S(c_\cdot),\; P(x_I,h_D) \xlongrightarrow{\frac{\kappa}{N}} S(c_Y),\;P(x_I,h_D)
\end{equation}
This is read as, \emph{any} surface coming into contact with an infectious
person with dirty hands becomes contaminated.
This happens at a rate $\kappa$ of contact with surfaces and proportionally to
the fraction of the population that is infectious with dirty hands.

Decontamination is a degradation rule,
\begin{equation}
  S(c_Y) \xlongrightarrow{\delta} S(c_N)
\end{equation}
where $\delta$ is the rate of surface cleaning or fomite degradation.
The interpretation of this is left open, it may simply be that the surface
contamination becomes incapable of transmitting the virus or that it is cleaned
every $\delta^{-1}$ time units.

The infection process is similar to the standard model, though it is a
consequence of interaction with contaminated surfaces rather than infectious
individuals,
\begin{equation}
  P(x_S),\;S(c_Y) \xlongrightarrow{\beta\frac{\kappa}{|S|}} P(x_E),\;S(c_Y)
\end{equation}
here, $\kappa$ is again the rate of contact with surfaces, and the rate of
infection is proportional to the fraction of contaminated surfaces.
The factor $\beta$ is, analogously to the standard model, the probability of
becoming infected upon exposure to a contaminated surface.

The rules for progression of the disease are exactly as for the model for mask
wearing in the previous section.

\begin{figure}[ht]
  \begin{subfigure}{0.48\textwidth}
    \resizebox{\textwidth}{!}{
      \begin{tikzpicture}
        \begin{axis}[
          cycle list name=color list, no markers,
          xmin=0, xmax=180, ymin=0, ymax=10000,
          legend style={at={(0.95,0.5)}, anchor=east, legend columns=1},
          ]
          \envelope{teal}{S}{\tiny $P(x_S)$}{data/nowash.tsv}
          \envelope{orange}{E}{\tiny $P(x_E)$}{data/nowash.tsv}
          \envelope{red}{I}{\tiny $P(x_I)$}{data/nowash.tsv}
          \envelope{brown}{R}{\tiny $P(x_R)$}{data/nowash.tsv}
          \envelope{blue}{C}{\tiny $S(C_Y)$}{data/nowash.tsv}
        \end{axis}
      \end{tikzpicture}
    }
    \caption{No hand-washing}
  \end{subfigure}
  \hfill
  \begin{subfigure}{0.48\textwidth}
    \resizebox{\textwidth}{!}{
      \begin{tikzpicture}
        \begin{axis}[
          cycle list name=color list, no markers,
          xmin=0, xmax=180, ymin=0, ymax=10000,
          legend style={at={(0.05,0.5)}, anchor=west, legend columns=1},
          ]
          \envelope{teal}{S}{\tiny $P(x_S)$}{data/fomites.tsv}
          \envelope{orange}{E}{\tiny $P(X_E)$}{data/fomites.tsv}
          \envelope{red}{I}{\tiny $P(x_I)$}{data/fomites.tsv}
          \envelope{brown}{R}{\tiny $P(x_R)$}{data/fomites.tsv}
          \envelope{blue}{C}{\tiny $S(c_Y)$}{data/fomites.tsv}
        \end{axis}
      \end{tikzpicture}
    }
    \caption{Hand washing}
  \end{subfigure}
  \caption{%
    The figure shows the results of a fomite model with and without hand
    washing. There is one shared surface for every four individuals. Individuals
    can have dirty, or clean hands. Hands are washed periodically (in this case
    eight times per day) and become dirty half-way between each washing.
    Infectious individuals with dirty hands contaminate surfaces.
    Susceptible individuals with dirty hands have a chance of being infected
    by contact with contaminated surfaces.
    Contaminated surfaces become spontaneously decontaminated after four hours,
    due to cleaning or degradation of the pathogen.
  }
  \label{fig:fomites}
\end{figure}
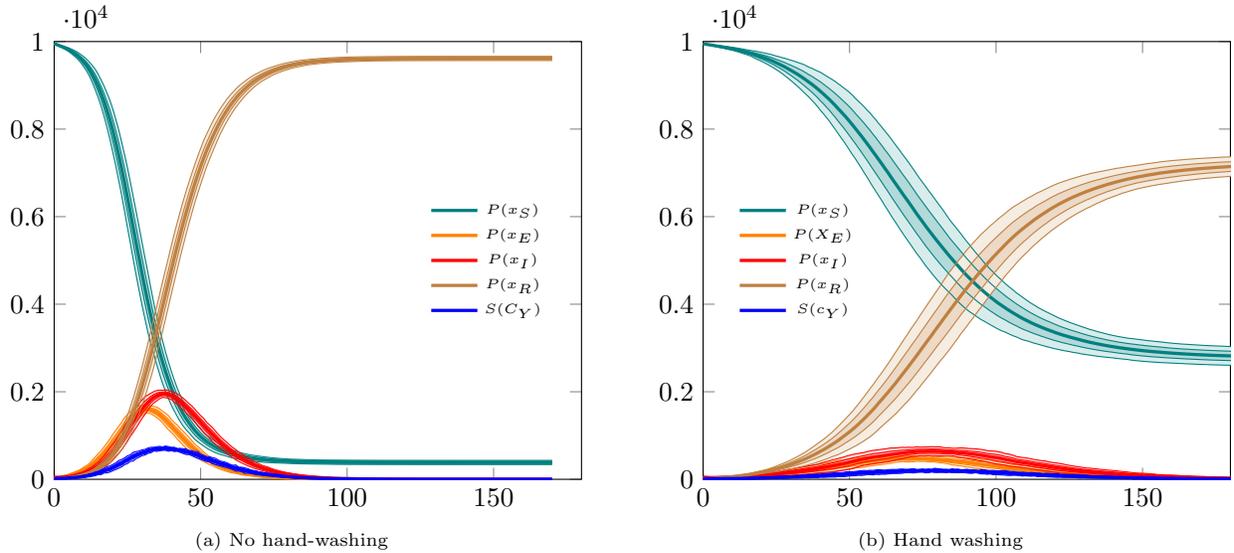
We can see in Figure~\ref{fig:fomites} the effect of hand washing with plausible
values for the rates.
The number of shared surfaces is taken to be equal to one quarter of the
population, and individuals come into contact with them twelve times per day.
For the case with hand washing, hands are washed relatively frequently, eight
times per day and become dirty at twice that rate.
Surfaces become decontaminated after four hours.
Hand washing results in a substantial reduction in the number of
contaminated surfaces which, in turn, causes a much smaller peak in the number
of infections and a lower number of cumulative infections.

Note also that exactly the same model, though likely with different rate
constants, is applicable to a scenario of transmission by aerosol.
This simply requires reinterpreting ``surface'' as ``indoor location'' since
these locations become contaminated through the presence of infected individuals
and the aerosols disperse after some time.
A slightly more sophisticated treatment that includes the effect of masks
analogously to the previous section is left as an exercise for the reader.

\subsection{Vectors}
Animate vectors of disease transmission such as mosquitoes may have a life-cycle
much shorter than the duration of a disease outbreak.
It would make sense to simply assume a constant population of vectors that
becomes carrier of disease and then ceases to be a carrier -- this may in fact
be the case in some circumstances.
However, for purposes of exposition, we choose to represent the birth and death
cycle of the vector explicitly.
Here, our agents will be,
\begin{align}
  P(x_u),&\;u\in \left\{ S, E, I, R \right\}\\
  V(x_v),&\;v\in \left\{ S, I \right\}
\end{align}
where the individuals simply have the states corresponding to disease
progression, and the vector may be susceptible or infectious.

We will just use a birth process depending only on the number of individual
vectors for simplicity and incorporate a vector control strategy,
\begin{equation}
  V() \xlongrightarrow{w k_b} V(),\; V(x_S)
\end{equation}
or in other words, a vector reproduces at rate $k_b$, and all offspring are
in the susceptible state regardless of the parent's infection status.
There is no transmission through reproduction among vectors, though that would
be trivial to do.
The factor, $w$, represents the destruction of breeding habitat and is allowed
to take on values in $[0,1]$.
If the vector is a mosquito, this could represent the fraction of the quantity of
standing water at the beginning of the simulation that is allowed to remain.

The death process is very simple and just happens at a constant rate, $k_d$,
\begin{equation}
  V() \xlongrightarrow{k_d} \emptyset
\end{equation}

In this model there are two kinds of infection process: people by vectors and
infection of vectors by people.
These happen with probabilities $\beta$ and $\beta^\prime$ respectively.
Let $M = |V|$ analogously to $N = |P|$, and we write for these rules,
\begin{align}
  P(x_S),\; V(x_I) &\xlongrightarrow{\beta\frac{\kappa}{N}} P(x_E),\; V(x_I)\\
  V(x_S),\; P(x_I) &\xlongrightarrow{\beta^\prime\frac{\kappa}{M}} V(x_I),\; P(x_I)
\end{align}
where $\kappa$ is the frequency at which a vector contacts (e.g. bites) the
host.

As before, disease progression for the host is unchanged from the above models.

\begin{figure}[ht]
  \begin{subfigure}{0.48\textwidth}
    \resizebox{\textwidth}{!}{
      \begin{tikzpicture}
        \begin{axis}[
          cycle list name=color list, no markers,
          xmin=0, xmax=360, ymin=0, ymax=10000,
          legend style={at={(0.05,0.5)}, anchor=west, legend columns=1},
          ]
          \envelope{teal}{S}{\tiny $P(x_S)$}{data/vector.tsv}
          \envelope{orange}{E}{\tiny $P(x_E)$}{data/vector.tsv}
          \envelope{red}{I}{\tiny $P(x_I)$}{data/vector.tsv}
          \envelope{brown}{R}{\tiny $P(x_R)$}{data/vector.tsv}
        \end{axis}
      \end{tikzpicture}
    }
    \caption{Host population}
  \end{subfigure}
  \hfill
  \begin{subfigure}{0.48\textwidth}
    \resizebox{\textwidth}{!}{
      \begin{tikzpicture}
        \begin{axis}[
          cycle list name=color list, no markers,
          xmin=0, xmax=360, ymin=0, ymax=50000,
          legend style={at={(0.05,0.5)}, anchor=west, legend columns=1},
          ]
          \envelope{teal}{Vs}{\tiny $V(x_S)$}{data/vector.tsv}
          \envelope{red}{Vi}{\tiny $V(x_I)$}{data/vector.tsv}
        \end{axis}
      \end{tikzpicture}
    }
    \caption{Vector population}
  \end{subfigure}
  \begin{subfigure}{0.48\textwidth}
    \resizebox{\textwidth}{!}{
      \begin{tikzpicture}
        \begin{axis}[
          cycle list name=color list, no markers,
          xmin=0, xmax=360, ymin=0, ymax=10000,
          legend style={at={(0.05,0.5)}, anchor=west, legend columns=1},
          ]
          \envelope{teal}{S}{\tiny $P(x_S)$}{data/habitat.tsv}
          \envelope{orange}{E}{\tiny $P(x_E)$}{data/habitat.tsv}
          \envelope{red}{I}{\tiny $P(x_I)$}{data/habitat.tsv}
          \envelope{brown}{R}{\tiny $P(x_R)$}{data/habitat.tsv}
        \end{axis}
      \end{tikzpicture}
    }
    \caption{Host population}
  \end{subfigure}
  \hfill
  \begin{subfigure}{0.48\textwidth}
    \resizebox{\textwidth}{!}{
      \begin{tikzpicture}
        \begin{axis}[
          cycle list name=color list, no markers,
          xmin=0, xmax=360, ymin=0, ymax=50000,
          legend style={at={(0.05,0.5)}, anchor=west, legend columns=1},
          ]
          \envelope{teal}{Vs}{\tiny $V(x_S)$}{data/habitat.tsv}
          \envelope{red}{Vi}{\tiny $V(x_I)$}{data/habitat.tsv}
        \end{axis}
      \end{tikzpicture}
    }
    \caption{Vector population}
  \end{subfigure}
  \caption{%
    Model of vector mediated disease transmission.
    Top row is the undisturbed system.
    On the left is the host population in its various states of disease
    progression and on the right is the vector population.
    Bottom row is the same system subjected to a perturbation: 10\% of the vectors'
    habitat is destroyed every 7 time periods.
  }
  \label{fig:vectors}
\end{figure}
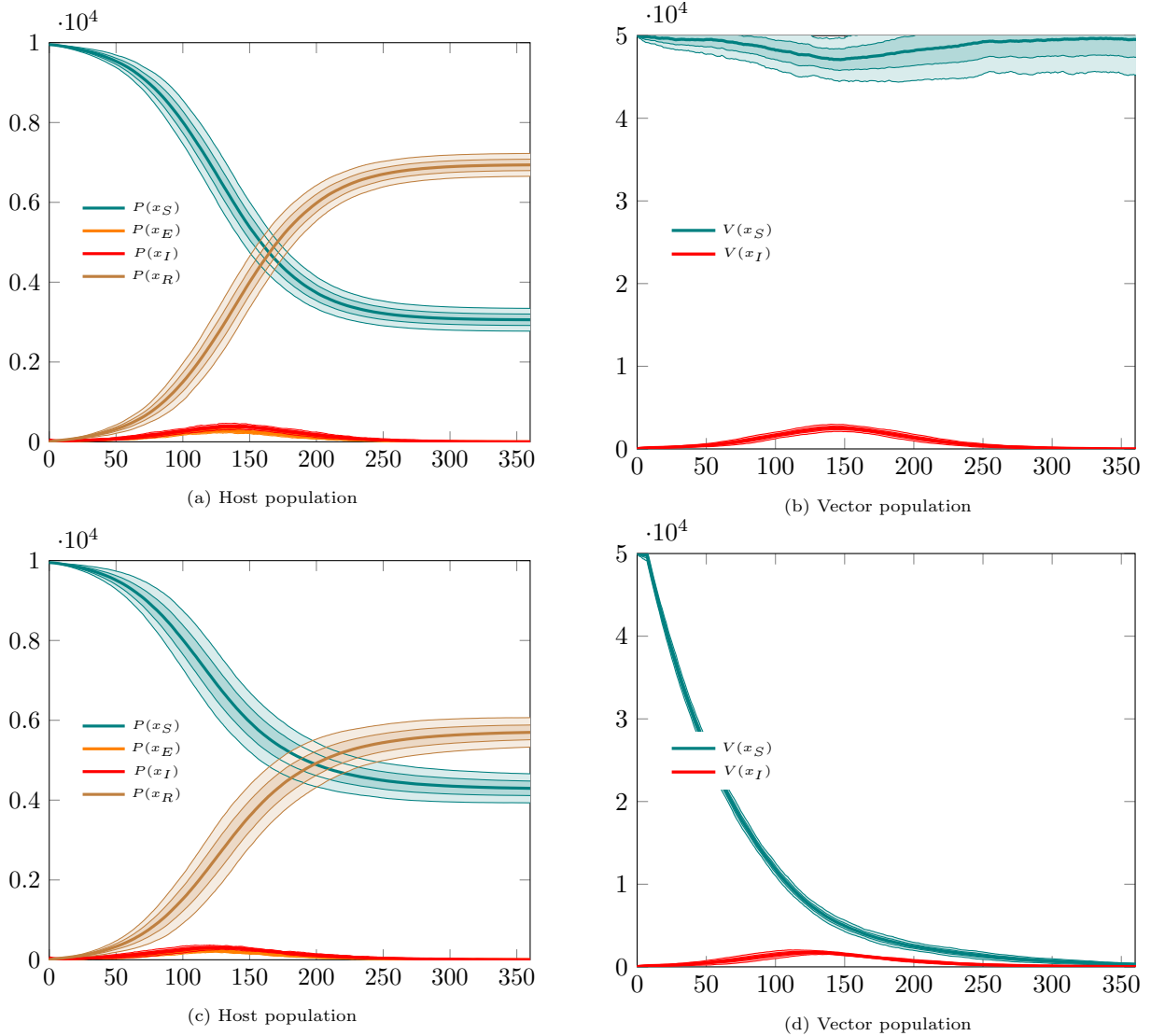
Figure~\ref{fig:vectors} shows the host and vector populations in this model
under two scenarios: undisturbed and where a 10\% of the vectors' habitat is
destroyed every 7 days.
The population of vectors starts out at five times that of the hosts and, in the
second scenario, precipitously declines, and brings the outbreak under control.

There are clearly elements of this scenario that could be modelled in more
detail.
An interesting observation, implicit in the birth-death process here, would be
the return of vectors to the susceptible state at some rate.
This could be made explicit with a $V(x_I)\rightarrow V(x_S)$ rule.
Offspring of vectors could, as mentioned above, inherit susceptibility or
infectiousness from the parent.
This scenario would then represent two coupled epidemiological models: an SEIR
model for the host and an SIS model for the vector.
These extensions, as before, are left as an exercise.

The code for this model is in Appendix~\ref{code:vectors}.

\subsection{Testing}
The presented rule-based approach can also allows to express a more sophisticated model of testing than is normally found.
In this model, tests, $T()$ are discrete units that are manufactured at a constant
rate $m$ and are consumed on use.
This has an advantage over a representation where tests are simply asserted to
be performed at some rate because if a test is not available, it cannot be
performed.
Explicitly representing tests as a participant in testing reflects important
considerations of manufacturing and the supply chain.
The manufacturing rule is simple,
\begin{equation}
  \emptyset \xlongrightarrow{m} T()
\end{equation}
and the tests have a characteristic recall (true positive rate), $r$ and
specificity (true negative rate), $s$.

Upon a positive test, an individual becomes isolated and can no longer infect
others.
In addition to the usual disease progression states and the quarantine state, we
also endow individuals with a ``testable'' state.
This last has no additional meaning and is true if and only if an individual is
infectious.
Its use is to allow for a more compact representation of the testing rules.
This is an example of where, far from adding complexity, the judicious addition
of states can actually simplify a model.
Our principal agent has the signature,
\begin{equation}
  P(x_u,t_v,q_w),\; u \in \{S, E, I, R\}, v \in \{Y, N\}, w \in \{Y, N\}
\end{equation}

Our progression and removal rules, while simple, are no longer the same as in
the previous models because they govern membership in the testable set. We
write,
\begin{align}
  P(x_E,t_N) &\xlongrightarrow{\alpha} P(x_I,t_Y)\\
  P(x_I,t_Y) &\xlongrightarrow{\gamma} P(x_R,t_N)
\end{align}
in other words, at the same instant that an individual becomes infectious, they
also become subject to correctly testing positive.
As they are removed through recovery or death, they are no longer subject to
correctly testing positive.

Infection is similar to a standard SEIR model with the caveat that it can only
take place among unconfined individuals,
\begin{equation}
  P(x_S,q_N),\; P(x_I,q_N) \xlongrightarrow{\beta\frac{c}{N}}
  P(x_E,q_N),\; P(x_I,q_N)
\end{equation}
Note that testability is not mentioned and isolation status is not changed.
This is exactly the standard infection rule applying only between the unconfined
subsets of $P(x_S)$ and $P(x_I)$.

There are four testing rules corresponding to the four possibilities of true
positives and negatives and false positives and negatives.
For realism, we suppose that there is a random testing rate, $\theta_0$,
for sampling the population, and a targetted testing rate, $\theta_I$, for
individuals that are infectious.
This is justified by the fact that infectious individuals are frequently
symptomatic, perhaps requiring medical care, and so they are specifically
tested.
Because infectious individuals may also be randomly testing, the effective
testing rate for them is,
\begin{equation}
  \theta = \theta_0 + \theta_I - \theta_0\theta_I
\end{equation}
where the third term on the right hand side corrects for double-counting as we
do not suppose that these individuals will be tested by both methods.

For present purposes, only unconfined individuals that participate in disease
propagation will be tested.
Our four testing rules are, therefore,
\begin{align}
  P(t_N,q_N),\; T() &\xlongrightarrow{\arate{s\theta_0}} P(t_N,q_N)\\
  P(t_N,q_N),\; T() &\xlongrightarrow{\arate{(1-s)\theta_0}} P(t_N,q_Y)\\
  P(t_Y,q_N),\; T() &\xlongrightarrow{\arate{r\theta}} P(t_N,q_Y)\\
  P(t_Y,q_N),\; T() &\xlongrightarrow{\arate{(1-r)\theta}} P(t_N,q_N)
\end{align}
These are, in order, true negatives, false negatives, true positives and false
positives.
Note that the test, $T()$, is consumed in this process and does not appear on
the right-hand side of any of the rules.
The reason for introducing the extra testable state is evident: it lets us write
these testing rules in terms of the relevant feature.
If we had not done so, it would have been necessary to write separate true and
false negative testing rules for each of $S, E, R$, resulting in eight rules in
total rather than four.

Finally, those individuals that became isolated when uninfected, or who have
recovered in isolation, exit to the unconfined state at a rate which we take
without loss of generality to be equal to the infectious period,
\begin{align}
  P(x_S,q_Y) &\xlongrightarrow{\gamma} P(x_S,q_N)\\
  P(x_R,q_Y) &\xlongrightarrow{\gamma} P(x_R,q_N)
\end{align}
which completes this model.
The corresponding code is reproduced in Appendix~\ref{code:testing}.

\begin{figure}[h!]
  \begin{subfigure}{0.48\textwidth}
    \resizebox{\textwidth}{!}{
      \begin{tikzpicture}
        \begin{axis}[
          cycle list name=color list, no markers,
          xmin=0, xmax=365, ymin=0, ymax=10000,
          legend style={at={(0.95,0.95)}, anchor=north east, legend columns=2},
          ]
          \envelope{teal}{Sn}{\tiny $P(x_S,q_N)$}{data/lowtest.tsv}
          \envelope{orange}{En}{\tiny $P(x_E,q_N)$}{data/lowtest.tsv}
          \envelope{red}{In}{\tiny $P(x_I,q_N)$}{data/lowtest.tsv}
          \envelope{brown}{Rn}{\tiny $P(x_Rq_N)$}{data/lowtest.tsv}
        \end{axis}
      \end{tikzpicture}
    }
    \caption{Unconfined population}
  \end{subfigure}
  \hfill
  \begin{subfigure}{0.48\textwidth}
    \resizebox{\textwidth}{!}{
      \begin{tikzpicture}
        \begin{axis}[
          cycle list name=color list, no markers,
          xmin=0, xmax=365, ymin=0, ymax=1000,
          legend style={at={(0.95,0.95)}, anchor=north east, legend columns=2},
          ]
          \envelope{teal}{Sy}{\tiny $P(x_S,q_Y)$}{data/lowtest.tsv}
          \envelope{orange}{Ey}{\tiny $P(x_E,q_Y)$}{data/lowtest.tsv}
          \envelope{red}{Iy}{\tiny $P(x_I,q_Y)$}{data/lowtest.tsv}
          \envelope{brown}{Ry}{\tiny $P(x_R,q_Y)$}{data/lowtest.tsv}
        \end{axis}
      \end{tikzpicture}
    }
    \caption{Isolated population}
  \end{subfigure}
  \begin{subfigure}{0.48\textwidth}
    \resizebox{\textwidth}{!}{
      \begin{tikzpicture}
        \begin{axis}[
          cycle list name=color list, no markers,
          xmin=0, xmax=365, ymin=0, ymax=10000,
          legend style={at={(0.95,0.95)}, anchor=north east, legend columns=2},
          ]
          \envelope{teal}{Sn}{\tiny $P(x_S,q_N)$}{data/testing.tsv}
          \envelope{orange}{En}{\tiny $P(x_E,q_N)$}{data/testing.tsv}
          \envelope{red}{In}{\tiny $P(x_I,q_N)$}{data/testing.tsv}
          \envelope{brown}{Rn}{\tiny $P(x_R,q_N)$}{data/testing.tsv}
        \end{axis}
      \end{tikzpicture}
    }
    \caption{Unconfined population}
  \end{subfigure}
  \hfill
  \begin{subfigure}{0.48\textwidth}
    \resizebox{\textwidth}{!}{
      \begin{tikzpicture}
        \begin{axis}[
          cycle list name=color list, no markers,
          xmin=0, xmax=365, ymin=0, ymax=1000,
          legend style={at={(0.95,0.95)}, anchor=north east, legend columns=2},
          ]
          \envelope{teal}{Sy}{\tiny $P(x_S,q_Y)$}{data/testing.tsv}
          \envelope{orange}{Ey}{\tiny $P(x_E,q_Y)$}{data/testing.tsv}
          \envelope{red}{Iy}{\tiny $P(x_I,q_Y)$}{data/testing.tsv}
          \envelope{brown}{Ry}{\tiny $P(x_R,q_Y)$}{data/testing.tsv}
        \end{axis}
      \end{tikzpicture}
    }
    \caption{Isolated population}
  \end{subfigure}
  \caption{
    Resource constrained testing leading to isolation.
    The top row is a circumstance where sufficient tests are manufactured to
    test 2.5\% of the population daily and the bottom row, 5\%.
    Tests are consumed on use, and have a recall (true positives rate)
    and specificity (true negative rate) of 80\%.
    A positive test leads immediately to isolation.
    On the left are the unconfined sub-populations and on the right are
    the isolated sub-populations.
    Infection only happens within the unconfined sub-populations.
  }
  \label{fig:testing}
\end{figure}
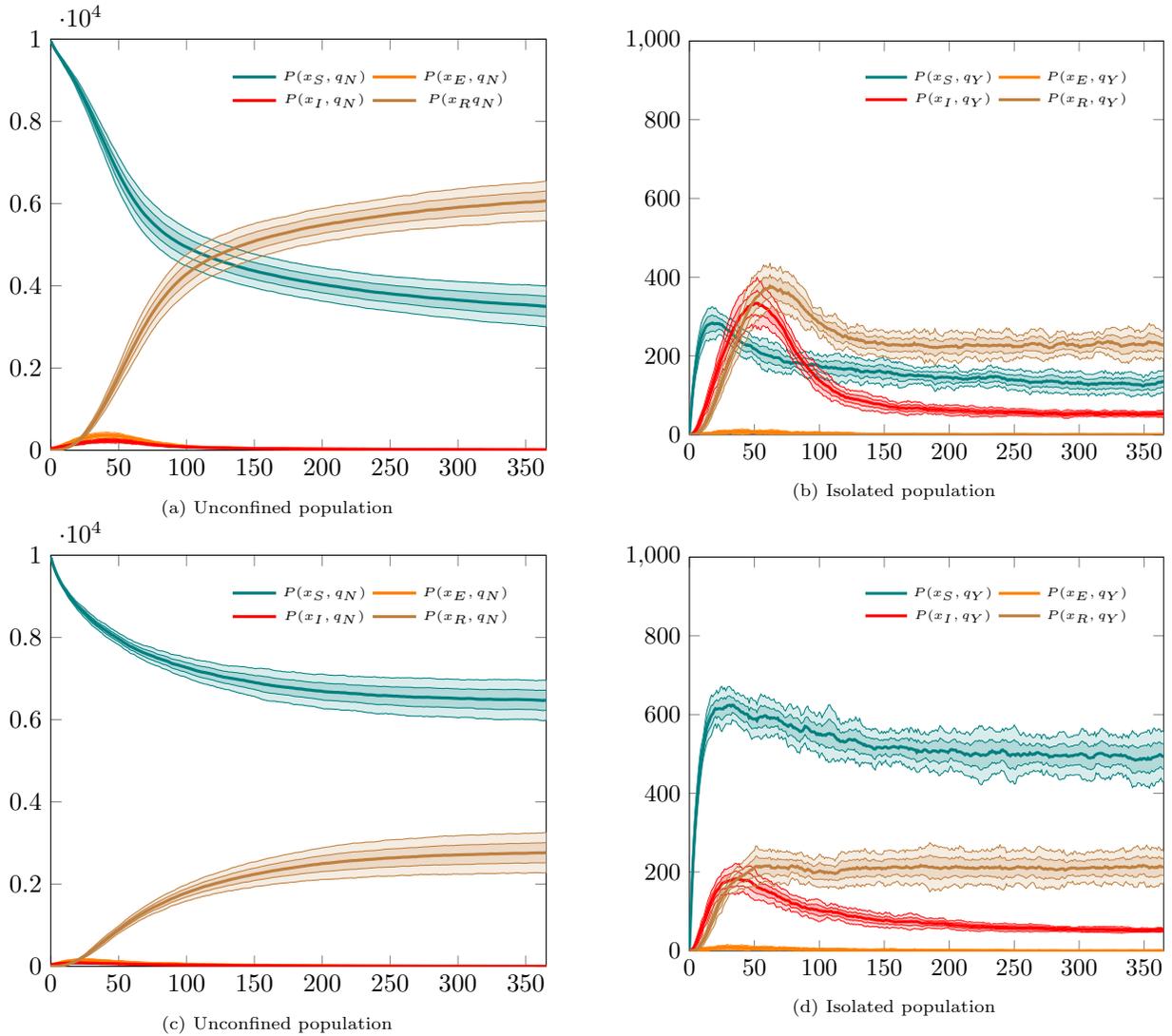
Figure~\ref{fig:testing} shows example trajectories of this system under
conditions of low and high production of tests.
For the top row, tests are manufactured at a rate sufficient to test 2.5\% of the population daily.
For the bottom row there are enough tests for 5\%.
An evident significant difference is the effect of false negatives with
increased testing.
In the bottom row, the majority of isolated individuals are, in fact,
susceptible or recovered and not infected.
The testing regime has a relatively high false positive rate of 20\% and because
the majority of the population is initially susceptible, there are more of them
isolated than the other population subsets.
As the outbreak progresses, more individuals are in the removed subset and they
become isolated in proportion to the fraction of the population that they make
up.
In the top row, infectious individuals initially dominate as there are
insufficient many tests to randomly sample the population.
These are also insufficiently many tests for isolation due to testing to contain
the outbreak, so, as it progresses, those who have recovered form the majority
of isolated individuals, again due to false positives in testing, simply because
they form the largest subset of the population.

\subsection{Tracing}
\label{sec:tti}

The final model in this paper which demonstrates the flexibility of the proposed rule-based approach is for contact tracing.

It builds upon the previous testing model because testing triggers contact
tracing.
It employs a slightly generalised version of the technique used in our previous
work~\cite{sturniolo_testing_2020} that functions as follows.
Suppose that each time contact with an infectious individual occurs, a trace is
left behind.
These traces follow individuals through the disease progression, and eventually
degrade.
We represent these traces as the agent,
\begin{equation}
  C(x_u),\; u \in \left\{ S, E, I, R \right\}
\end{equation}
and we use the same agents $T$ and $P$ from the previous model.

The progression rules for individuals are likewise the same, and to then we add
straightforward equivalents for the traces, along with a degradation rule,
\begin{align}
  C(x_E) &\xlongrightarrow{\alpha} C(x_I)\\
  C(x_I) &\xlongrightarrow{\gamma} C(x_R)\\
  C() &\xlongrightarrow{\gamma} \emptyset
\end{align}

Contact is, however, now more complex as we need rules for all contacts with
infectious individuals, not only those that result in infection.
All of these contacts produce traces, but only those which result in infection
change the state of the individual contact,
\begin{align}
  P(x_S,q_N),\; P(x_I,q_N) &\xlongrightarrow{\arate{\frac{(1- \beta) c}{N}}} P(x_S,q_N),\; P(x_I,q_N),\; C(x_S)\\
  P(x_S,q_N),\; P(x_I,q_N) &\xlongrightarrow{\arate{\frac{\beta c}{N}}} P(x_E,q_N),\; P(x_I,q_N),\; C(x_E)\\
  P(x_E,q_N),\; P(x_I,q_N) &\xlongrightarrow{\arate{\frac{c}{N}}} P(x_E,q_N),\; P(x_I,q_N),\; C(x_E)\\
  P(x_I,q_N),\; P(x_I,q_N) &\xlongrightarrow{\arate{\frac{c}{N}}} P(x_I,q_N),\; P(x_I,q_N),\; C(x_I)\\
  P(x_R,q_N),\; P(x_I,q_N) &\xlongrightarrow{\arate{\frac{c}{N}}} P(x_R,q_N),\; P(x_I,q_N),\; C(x_R)
\end{align}

Tracing is an operation that consumes a trace.
Individuals may be traced whether or not they are isolated, and are traced in
proportion to the fraction of the isolated population: those who have been
isolated due to a true or false positive test have their contacts traced.
The tracing rules, therefore, are,
\begin{equation}
  P(x_u),\; P(q_Y),\; C(x_S) \xlongrightarrow{\frac{\eta\theta}{N}} P(x_u,q_Y),\; P(q_Y)
\end{equation}
for $u \in \{S,E,I,R\}$, and where $\eta$ is the tracing efficiency, or the number
of contacts per unit time that are expected to be traced.

Note that this formulation of tracing differs from that
of our previous work~\cite{sturniolo_testing_2020} in two main respects.
First, tracing is somewhat recursive: becoming isolated due to tracing also
causes contacts to be traced.
If recursive tracing is not required, it is sufficient to add a state that
records test results.
Secondly, here, the rate of being traced is proportional to the number of
infectious contacts one has experienced in the time window before the traces
degrade.
If tracing happens at a constant rate per infectious individual, then having had
contact with two such individuals should result in being traced more quickly.
There is no mechanism, however, to distinguish between multiple contacts with
the same infectious individual and contacts with multiple infectious
individuals.
In our previous work, tracing happens in proportion to the
likelihood of having had \emph{at least one} infectious contact which may tend
to underestimate the influence of contact tracing on containment of outbreaks.
The model given here is more eager and this may lead to overestimation the effect of
tracing.
It is not obvious which model most closely resembles the reality of contact
tracing.

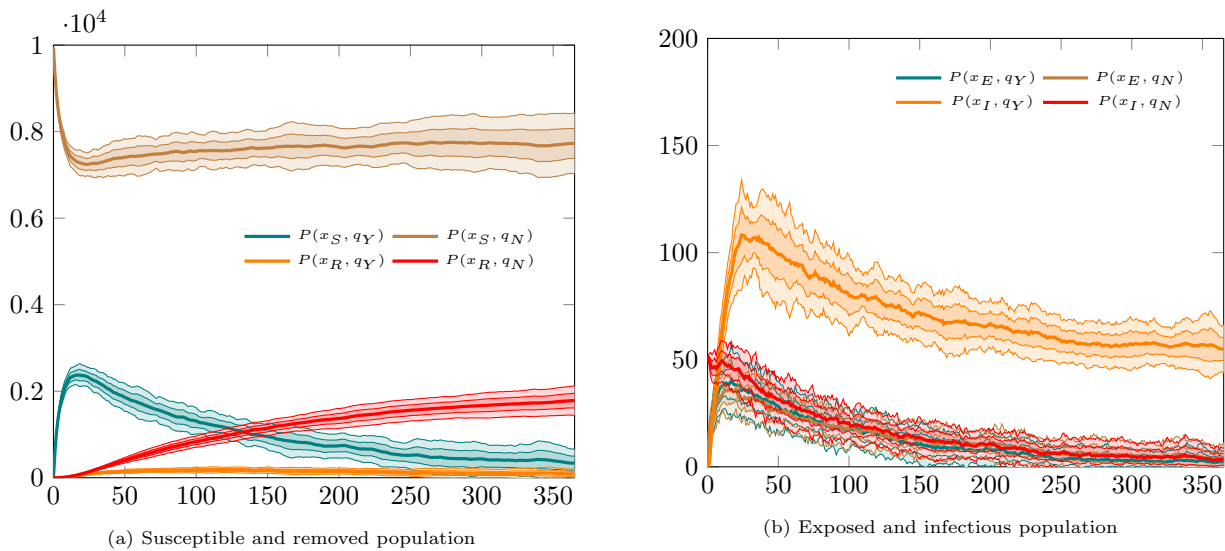
\begin{figure}[ht]
  \begin{subfigure}{0.48\textwidth}
    \resizebox{\textwidth}{!}{
      \begin{tikzpicture}
        \begin{axis}[
          cycle list name=color list, no markers,
          xmin=0, xmax=365, ymin=0, ymax=10000,
          legend style={at={(0.95,0.6)}, anchor=north east, legend columns=2},
          ]
          \envelope{teal}{Sy}{\tiny $P(x_S,q_Y)$}{data/tracing.tsv}
          \envelope{brown}{Sn}{\tiny $P(x_S,q_N)$}{data/tracing.tsv}
          \envelope{orange}{Ry}{\tiny $P(x_R,q_Y)$}{data/tracing.tsv}
          \envelope{red}{Rn}{\tiny $P(x_R,q_N)$}{data/tracing.tsv}
        \end{axis}
      \end{tikzpicture}
    }
    \caption{Susceptible and removed population}
  \end{subfigure}
  \hfill
  \begin{subfigure}{0.48\textwidth}
    \resizebox{\textwidth}{!}{
      \begin{tikzpicture}
        \begin{axis}[
          cycle list name=color list, no markers,
          xmin=0, xmax=365, ymin=0, ymax=200,
          legend style={at={(0.95,0.95)}, anchor=north east, legend columns=2},
          ]
          \envelope{teal}{Ey}{\tiny $P(x_E,q_Y)$}{data/tracing.tsv}
          \envelope{brown}{En}{\tiny $P(x_E,q_N)$}{data/tracing.tsv}
          \envelope{orange}{Iy}{\tiny $P(x_I,q_Y)$}{data/tracing.tsv}
          \envelope{red}{In}{\tiny $P(x_I,q_N)$}{data/tracing.tsv}
        \end{axis}
      \end{tikzpicture}
    }
    \caption{Exposed and infectious population}
  \end{subfigure}
  \caption{
    Isolation through testing and contact tracing.
    Note the different layout from previous figures in order to obtain better
    agreement of scale.
    On the left are susceptible and removed individuals, both isolated and not.
    On the right are infectious and exposed individuals, both isolated and not.
    The parameters are as with those for the scenario with testing alone,
    identical background rates and targeted rates of testing, and
    identical 80\% accuracy.
    Here, manufacturing is even more constrained: only 100 tests/day are
    produced.
    Contact tracing is performed efficiently with 90\% of contacts traced in 48
    hours. 
  }
  \label{fig:tracing}
\end{figure}
Figure~\ref{fig:tracing} shows this model performing under less optimal testing
conditions than previously.
That is, the tests are identically 80\% accurate and the aspirational sampling
and targetting testing rates are the same.
The testing rates are aspirational because manufacturing is even more
constrained: only 100 tests produced per day.
This poor provision of tests is supplemented with a good contact tracing regime
as described above with $\eta = 0.45$ meaning that 90\% of contacts are traced,
on average, within two days.
A striking feature is the large and slowly degrading number of susceptible
individuals that are isolated due to contact tracing.
This is a consequence of the fact that it is far less likely to become infected
due to contact with an infectious individual than to escape infection.
These contacts are nevertheless traced, resulting in many susceptible
individuals becoming isolated.
This number is sufficiently large that it appreciably depletes the susceptible
pool, rapidly slowing propagation of the disease.

\subsection{Schools}

Our next example, reproduced in code in Appendix~\ref{code:school} and shows how the proposed rule-based approach can express two coupled subpopulations: adults and children.
The background environment is a well-mixed epidemic model such as we have seen
above, with the usual progression rules and an infection rule attenuated with
contact restrictions.
Against this background, some structure is added: families.
A family may consist of up to two children and up to two adults.
The infection is transmitted much more easily within families; family members
are in frequent close contact with one another.
We also allow children to go to school, a second well-mixed environment.
Though children spend only part of their time at school, contact with other
children is much more frequent.
Let us see how to represent this rather complex situation as a small number of
rules.

We begin by defining the primary agent.
It has the same infection states as above, as well as an internal state to
identify as either a child or an adult.
Additionally, there are three binding sites that permit the formation of
child-parent or parent-parent bonds,
\begin{equation}
  P(x_u, a_v, \mathbf{e}^{i,j,k}),\: u \in \left\{ S, E, I, R \right\},
  v \in \left\{ A, C \right\}, i, j, k \in \mathbb{N^+}
\end{equation}

We use fast binding rules to bind pairs of adults at the beginning of the
simulation,
\begin{equation}
  P(a_A, e_3^\cdot), P(a_A, e_3^\cdot) \xlongrightarrow{\infty}   P(a_A, e_3^1), P(a_A, e_3^1)
\end{equation}
and then we assign bind to pairs of adults,
\begin{align}
  P(a_C, e_1^\cdot, e_2^\cdot), P(a_A, e_1^\cdot, e_3^3), P(a_A, e_2^\cdot, e_3^3)
  \xlongrightarrow{\infty}\\\nonumber
  P(a_C, e_1^1, e_2^2), P(a_A, e_1^1, e_3^3), P(a_A, e_2^2, e_3^3)
\end{align}
These three rules are sufficient to generate a small variety of family motifs: single individuals, childless couples, and couples with one or two children.

The main feature of these two rules is that the motifs that they produce are bounded in size.
An alternative formulation would first associate children with adults and then preferentially associate the parents of the same children.
This would allow for families with two children and three parents, or indeed arbitrarily large families in different configurations.
Such a generative rule-set could be,
\begin{align}
    P(a_C, e_1^\cdot), P(a_A, e_1^\cdot) &\xlongrightarrow{\infty} P(a_C, e_1^\cdot), P(a_A, e_1^\cdot)\\
    P(a_C, e_2^\cdot), P(a_A, e_2^\cdot) &\xlongrightarrow{\infty} P(a_C, e_2^\cdot), P(a_A, e_2^\cdot)\\
    P(a_C, e_1^1, e_2^2), P(a_A, e_1^1, e_3^\cdot), P(a_A, e_2^2, e_3^\cdot) &\xlongrightarrow{\infty}\\\nonumber
    P(a_c, e_1^1, e_2^2), P(a_A, e_1^1, e_3^3), P(a_A, e_2^2, e_3^3)
\end{align}
and may reflect human society more accurately.
The distribution of motifs could be adjusted by using different large but finite rates rather than $\infty$.

The simulation substrate being a regular SEIR model, we have standard infection
and progression rules that will be familiar from the foregoing sections,
\begin{align}
  P(x_S), P(x_I) &\xlongrightarrow{\arate{\ell\beta\frac{c}{N}}} P(x_E), P(x_I)\\
  P(x_E) &\xlongrightarrow{\alpha} P(x_I)\\
  P(x_I) &\xlongrightarrow{\gamma} P(x_R)
\end{align}
where $\ell$ is the factor by which lockdown distancing measures attenuate the
normal propagation of the disease.

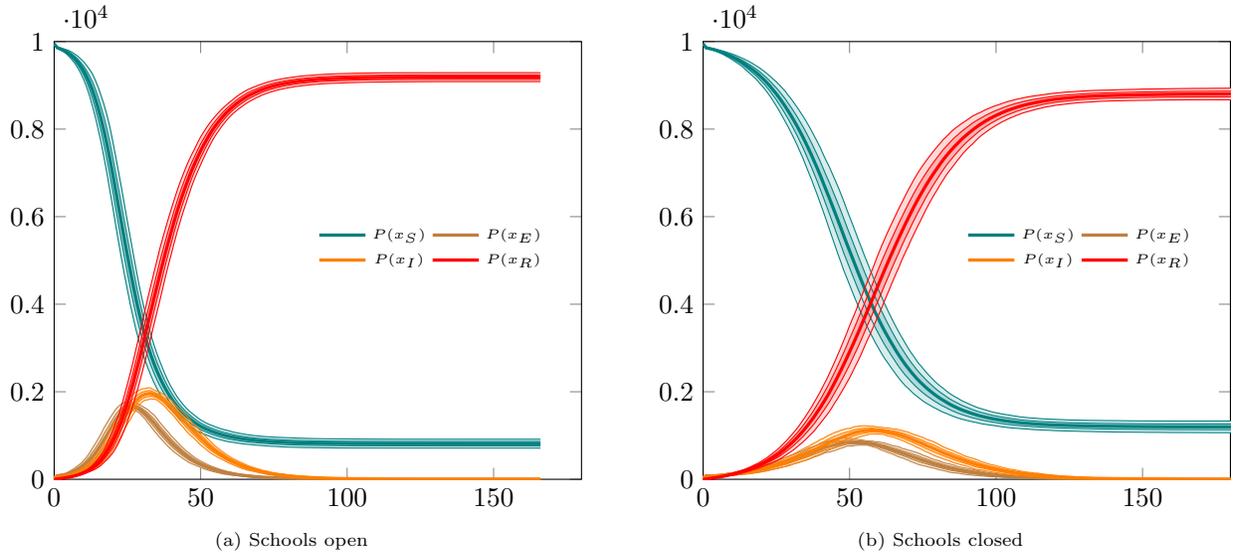
\begin{figure}[ht]
  \begin{subfigure}{0.48\textwidth}
    \resizebox{\textwidth}{!}{
      \begin{tikzpicture}
        \begin{axis}[
          cycle list name=color list, no markers,
          xmin=0, xmax=180, ymin=0, ymax=10000,
          legend style={at={(0.95,0.6)}, anchor=north east, legend columns=2},
          ]
          \envelope{teal}{S}{\tiny $P(x_S)$}{data/school.tsv}
          \envelope{brown}{E}{\tiny $P(x_E)$}{data/school.tsv}
          \envelope{orange}{I}{\tiny $P(x_I)$}{data/school.tsv}
          \envelope{red}{R}{\tiny $P(x_R)$}{data/school.tsv}
        \end{axis}
      \end{tikzpicture}
    }
    \caption{Schools open}
  \end{subfigure}
  \hfill
  \begin{subfigure}{0.48\textwidth}
    \resizebox{\textwidth}{!}{
      \begin{tikzpicture}
        \begin{axis}[
          cycle list name=color list, no markers,
          xmin=0, xmax=180, ymin=0, ymax=10000,
          legend style={at={(0.95,0.6)}, anchor=north east, legend columns=2},
          ]
          \envelope{teal}{S}{\tiny $P(x_S)$}{data/noschool.tsv}
          \envelope{brown}{E}{\tiny $P(x_E)$}{data/noschool.tsv}
          \envelope{orange}{I}{\tiny $P(x_I)$}{data/noschool.tsv}
          \envelope{red}{R}{\tiny $P(x_R)$}{data/noschool.tsv}
        \end{axis}
      \end{tikzpicture}
    }
    \caption{Schools closed}
  \end{subfigure}
  \caption{The influence of schools on disease propagation. On the left is the
    trajectory of an epidemic with schools as described in the text. One quarter
    of the population consists of children and distancing measures are generally
    in effect, except that the children go to school where they interact at a
    high rate. On the right, schools are closed, $s = 0$. 
    Though the asymptotic distribution is similar, schools promote infection
    propagation through the otherwise distanced population.}
  \label{fig:schools}
\end{figure}

The children in this simulation spend some time in school, a fraction, $s \in [0,1]$, of
their waking day.
While at school, they have contact with other children at an accelerated rate,
$\kappa > 1$.
This phenomenon will be familiar to any parent of a school-aged child.
School is represented simply as a second mass-action infection rule that applies
only to children.
\begin{equation}
  P(a_C, x_S), P(a_C, x_I) \xlongrightarrow{\arate{s\kappa\beta\frac{c}{N}}}
  P(a_C, x_E), P(a_C, x_I)
\end{equation}

Finally, infection does not spread within families with the same dynamic as in
the general population.
Family members are much more likely to pass the infection to each other.
We use three rules for this: from the child to each parent in proportion to the
time they spend away from school, and between parents,
\begin{align}
  P(x_S, e_1^1), P(x_I, e_1^1) &\xlongrightarrow{\arate{(1-s)\frac{\beta}{d}}} P(x_E, e_1^1), P(x_I, e_1^1)\\
  P(x_S, e_2^2), P(x_I, e_2^2) &\xlongrightarrow{\arate{(1-s)\frac{\beta}{d}}} P(x_E, e_2^2), P(x_I, e_2^2)\\
  P(x_S, e_3^3), P(x_I, e_3^3) &\xlongrightarrow{\arate{\frac{\beta}{d}}} P(x_E, e_3^3), P(x_I, e_3^3)
\end{align}

\begin{figure}[ht]
  \begin{subfigure}{0.48\textwidth}
    \resizebox{\textwidth}{!}{
      \begin{tikzpicture}
        \begin{axis}[
          cycle list name=color list, no markers,
          xmin=0, xmax=180, ymin=0, ymax=10000,
          legend style={at={(0.95,0.95)}, anchor=north east, legend columns=2},
          ]
          \envelope{teal}{Sa}{\tiny $P(x_S,a_A)$}{data/school.tsv}
          \envelope{brown}{Sc}{\tiny $P(x_S,a_C)$}{data/school.tsv}
          \envelope{orange}{Ra}{\tiny $P(x_R,a_A)$}{data/school.tsv}
          \envelope{red}{Rc}{\tiny $P(x_R,a_C)$}{data/school.tsv}
        \end{axis}
      \end{tikzpicture}
    }
    \caption{Schools open, $S$ \& $R$}
  \end{subfigure}
  \hfill
  \begin{subfigure}{0.48\textwidth}
    \resizebox{\textwidth}{!}{
      \begin{tikzpicture}
        \begin{axis}[
          cycle list name=color list, no markers,
          xmin=0, xmax=180, ymin=0, ymax=2000,
          legend style={at={(0.95,0.95)}, anchor=north east, legend columns=2},
          ]
          \envelope{teal}{Ea}{\tiny $P(x_E,a_A)$}{data/school.tsv}
          \envelope{brown}{Ec}{\tiny $P(x_E,a_C)$}{data/school.tsv}
          \envelope{orange}{Ia}{\tiny $P(x_I,a_A)$}{data/school.tsv}
          \envelope{red}{Ic}{\tiny $P(x_I,a_C)$}{data/school.tsv}
        \end{axis}
      \end{tikzpicture}
    }
    \caption{Schools open, $E$ \& $I$}
    \label{fig:openei}
  \end{subfigure}
  \begin{subfigure}{0.48\textwidth}
    \resizebox{\textwidth}{!}{
      \begin{tikzpicture}
        \begin{axis}[
          cycle list name=color list, no markers,
          xmin=0, xmax=180, ymin=0, ymax=10000,
          legend style={at={(0.95,0.95)}, anchor=north east, legend columns=2},
          ]
          \envelope{teal}{Sa}{\tiny $P(x_S,a_A)$}{data/noschool.tsv}
          \envelope{brown}{Sc}{\tiny $P(x_S,a_C)$}{data/noschool.tsv}
          \envelope{orange}{Ra}{\tiny $P(x_R,a_A)$}{data/noschool.tsv}
          \envelope{red}{Rc}{\tiny $P(x_R,a_C)$}{data/noschool.tsv}
        \end{axis}
      \end{tikzpicture}
    }
    \caption{Schools closed, $S$ \& $R$}
  \end{subfigure}
  \hfill
  \begin{subfigure}{0.48\textwidth}
    \resizebox{\textwidth}{!}{
      \begin{tikzpicture}
        \begin{axis}[
          cycle list name=color list, no markers,
          xmin=0, xmax=180, ymin=0, ymax=2000,
          legend style={at={(0.95,0.95)}, anchor=north east, legend columns=2},
          ]
          \envelope{teal}{Ea}{\tiny $P(x_E,a_A)$}{data/noschool.tsv}
          \envelope{brown}{Ec}{\tiny $P(x_E,a_C)$}{data/noschool.tsv}
          \envelope{orange}{Ia}{\tiny $P(x_I,a_A)$}{data/noschool.tsv}
          \envelope{red}{Ic}{\tiny $P(x_I,a_C)$}{data/noschool.tsv}
        \end{axis}
      \end{tikzpicture}
    }
    \caption{Schools closed, $E$ \& $I$}
    \label{fig:closedei}
  \end{subfigure}
  \caption{Schools are an accelerant. This is a more detailed view of the same
    scenarios of Figure~\ref{fig:schools} with the adult and child population
    shown separately. Note in particular the curve for $P(x_I,a_C)$ clearly leading
    the curve for $P(x_I,a_P)$ in Figure~\ref{fig:openei}. This is different from
    Figure~\ref{fig:closedei} where the corresponding peaks occur at similar
    times. This shows that children attending schools are agents of
    infection of the wider population.}
  \label{fig:accelerant}
\end{figure}
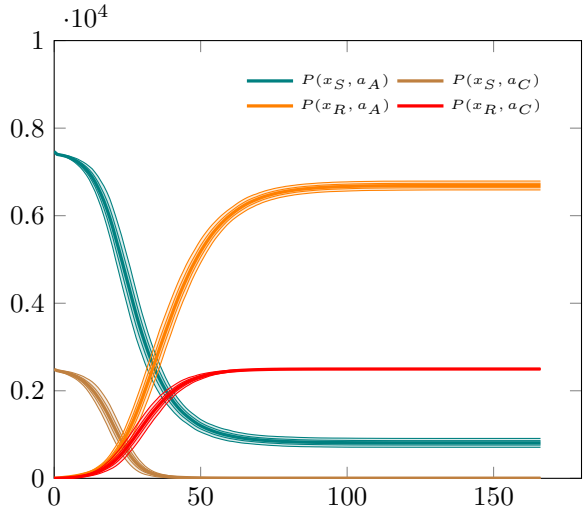
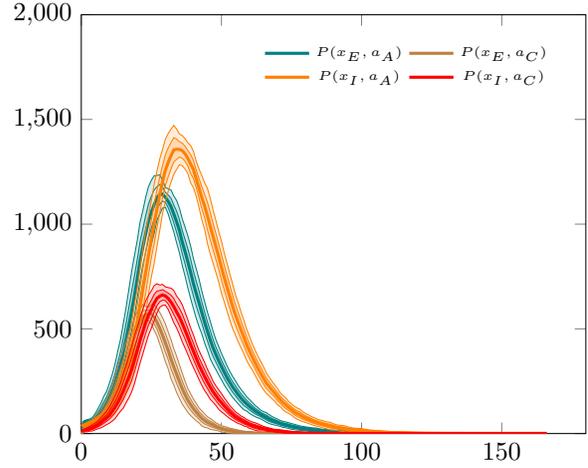
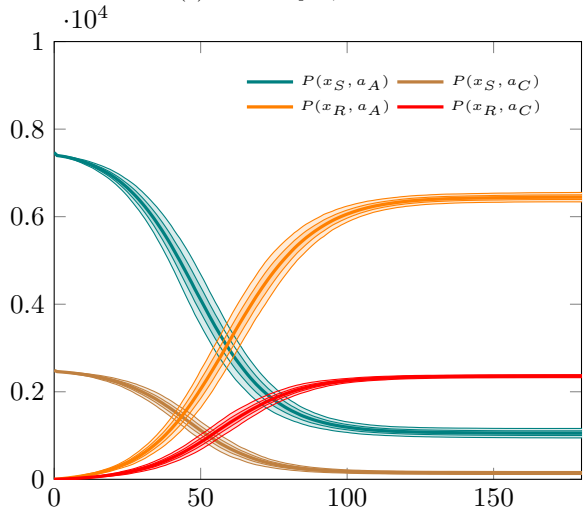
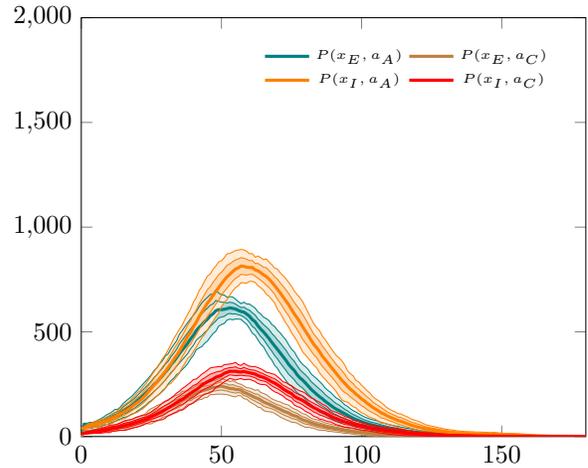

That is the entire model: two compartmental-style infection processes coupled
with a network epidemic model, expressed succinctly in 9 rules.
The results at the level of the population are shown in
Figure~\ref{fig:schools}.
The overall effect of schools being open is clear: the peak in infections is
much larger, and the outbreak progresses more rapidly though there is little
change in the cumulative infections (equivalent to $P(x_R)$). 
The underlying mechanism is visible in Figure~\ref{fig:accelerant} where the
adult and child subpopulations are presented separately.
In particular, Figure~\ref{fig:openei} shows the curve for infectious children,
$P(x_I,a_C)$, significantly leading that for infectious adults, $P(x_I,a_A)$.
Schools, modelled as we have done here, are an accelerant of the outbreak.
Of course, this is true in this case by construction: we supposed that a
sub-population spends some time in circumstances where contact happens at a
greater rate than in the general population.
However such a representation would seem to correspond reasonably faithfully to
reality.

\subsection{Gatherings}
\label{sec:gatherings}

Our final example is one kind of superspreading event.
There are several scenarios in which such events can occur, driven by
biological, behavioural, environmental factors or indeed
happenstance~\cite{althouse_stochasticity_2020}.
This example is of the behavioural kind.
The agents in this case are placed on a spectrum from ``loner'' to
``socialite''.
The difference is the propensity to participate in ``gatherings'' which are
daily events where the contact rate is much higher than normal.
Whereas in previous examples, the disease was parametrised to have an
infectiousness ($\beta$) comparable to what we expect from the 2019 novel
coronavirus, here we use an contagion that is only half as infectious.

The agent in this simulation is declared as,
\begin{equation}
  P(x_u, g_v, c_n)\:u\in\left\{ S,E,I,R \right\},\;v\in\left\{ Y,N \right\},\;c\in\left\{ 1,\ldots,10 \right\}
\end{equation}
The $g$ site indicates whether the individual is participating in a gathering,
and $c$ is an integer scale from 1 to 10 of how social that individual is
implemented using a counter.
Beginning with some housekeeping, as the internal state of $c$ will be
initialised to zero, we very rapidly assign individuals uniformly to the social
scale,
\begin{equation}
  P(c_0) \xlongrightarrow{\infty} P(c_n)
\end{equation}
for each $n$.

Progression of the disease are the standard simple rules,
$P(x_E) \xrightarrow{\alpha} P(x_I)$ and $P(x_I) \xrightarrow{\gamma} P(x_R)$,
and we have a pair of infection rules,
\begin{align}
  P(x_S, g_N), P(x_I, g_N) &\xlongrightarrow{\arate{\beta \frac{c_n}{|P(g_N)|}}} P(x_E, g_N), P(x_I, g_N)\\
  P(x_S, g_Y), P(x_I, g_Y) &\xlongrightarrow{\arate{\beta \frac{c_g}{|P(g_Y)|}}} P(x_E, g_Y), P(x_I, g_Y)
\end{align}
These are similar to the standard infection rule, though the rates are
different.
First $c_g \gg c_n$, the contact rate at gatherings is much higher than usual,
and the normalisation constant is not the entire population but only those that
are gathering, or not as appropriate.
This represents a true partition of the population into those that are gathering
and interacting only with one another, and those that are not.
There is no interaction between gatherings and the rest of the population while
the gathering is taking place -- those social creatures that gather become
infected and take the disease home.

The remaining two rules describe joining and leaving a gathering,
\begin{align}
  \label{eq:gather}
  P(g_N, c_n) &\xlongrightarrow{\textfrak{g}k_g\frac{n}{10}} P(g_Y, c_n)\\
  P(g_Y) &\xlongrightarrow{k_n} P(g_N)
\end{align}
If $k_g$ is the maximum rate of joining gatherings, the rate at which
Equation~\ref{eq:gather} occurs is scaled down according to the social
predisposition of the individual.
In the simulation $k_g$ is chosen such that the most social individuals are
expected to gather once per day and $k_n$ such that they are expected to leave a
gathering after an hour.
The $\textfrak{g}$ is a binary parameter indicating whether a gathering is taking
place or not.
This is set and cleared with a pair of perturbations.
It is set to 1 once per day and set to 0 after six hours.

\begin{figure}[ht]
  \centering
  \begin{tikzpicture}
    \begin{axis}[
      cycle list name=color list, no markers,
      xmin=0, xmax=14, ymin=0,
      legend style={at={(0.95,0.95)}, anchor=north east, legend columns=2},
      ]
      \envelope{teal}{G}{\tiny $P(g_Y)$}{supercut.tsv}
    \end{axis}
  \end{tikzpicture}
  \caption{Gatherings. This figure shows a two week time period with regular gatherings. Individuals spend about an hour at a gathering, and at most these consist of about 250 individuals (2.5\% of the population).}
  \label{fig:gathering}
\end{figure}
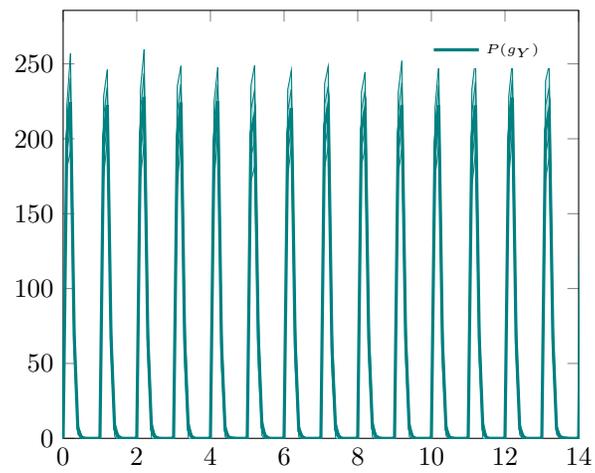
Figure~\ref{fig:gathering} shows daily gatherings in a two week time period.
Individuals spend an hour, on average, at a gathering, and at their peak, gatherings consist of about 2.5\% of the population, though for a short time.
The contact rate within a gathering is 10$\times$ the normal rate.
This increase corresponds to a modest increase in the average contact rate of the most social individuals of a factor of 
$\frac{23}{24} + 10\frac{1}{24} = 1.375$.
Only 10\% of the population is that gregarious; on average these gatherings increase the contact rate by a factor of only 1.17. 

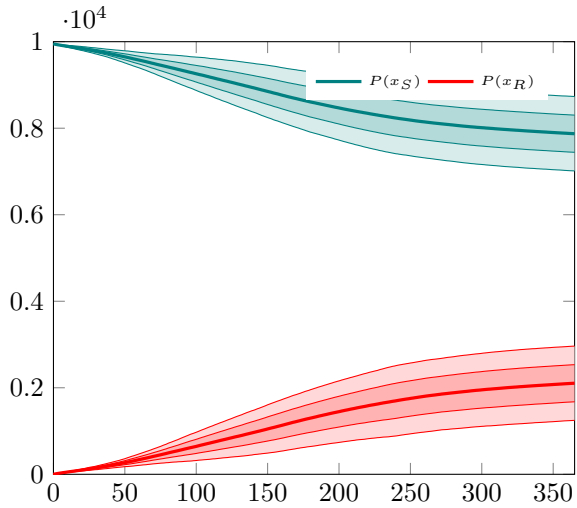
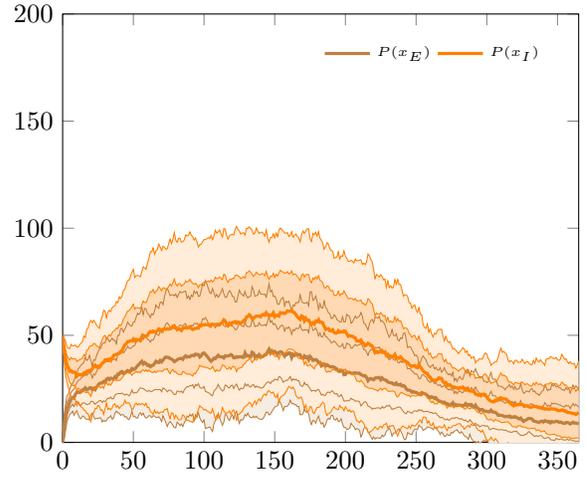
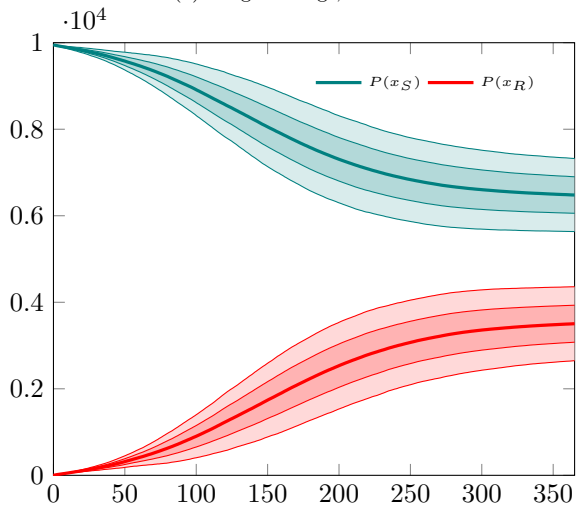
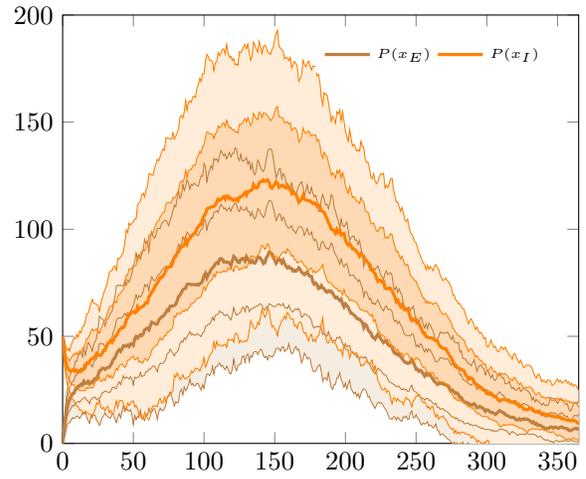
\begin{figure}[ht]
  \begin{subfigure}{0.48\textwidth}
    \resizebox{\textwidth}{!}{
      \begin{tikzpicture}
        \begin{axis}[
          cycle list name=color list, no markers,
          xmin=0, xmax=365, ymin=0, ymax=10000,
          legend style={at={(0.95,0.95)}, anchor=north east, legend columns=2},
          ]
          \envelope{teal}{S}{\tiny $P(x_S)$}{data/nosuper.tsv}
          \envelope{red}{R}{\tiny $P(x_R)$}{data/nosuper.tsv}
        \end{axis}
      \end{tikzpicture}
    }
    \caption{No gatherings, $S$ \& $R$}
  \end{subfigure}
  \hfill
  \begin{subfigure}{0.48\textwidth}
    \resizebox{\textwidth}{!}{
      \begin{tikzpicture}
        \begin{axis}[
          cycle list name=color list, no markers,
          xmin=0, xmax=365, ymin=0, ymax=200,
          legend style={at={(0.95,0.95)}, anchor=north east, legend columns=2},
          ]
          \envelope{brown}{E}{\tiny $P(x_E)$}{data/nosuper.tsv}
          \envelope{orange}{I}{\tiny $P(x_I)$}{data/nosuper.tsv}
        \end{axis}
      \end{tikzpicture}
    }
    \caption{No gatherings, $E$ \& $I$}
  \end{subfigure}
  \par%
  \begin{subfigure}{0.48\textwidth}
    \resizebox{\textwidth}{!}{
      \begin{tikzpicture}
        \begin{axis}[
          cycle list name=color list, no markers,
          xmin=0, xmax=365, ymin=0, ymax=10000,
          legend style={at={(0.95,0.95)}, anchor=north east, legend columns=2},
          ]
          \envelope{teal}{S}{\tiny $P(x_S)$}{supersift.tsv}
          \envelope{red}{R}{\tiny $P(x_R)$}{supersift.tsv}
        \end{axis}
      \end{tikzpicture}
    }
    \caption{Gatherings, $S$ \& $R$}
  \end{subfigure}
  \hfill
  \begin{subfigure}{0.48\textwidth}
    \resizebox{\textwidth}{!}{
      \begin{tikzpicture}
        \begin{axis}[
          cycle list name=color list, no markers,
          xmin=0, xmax=365, ymin=0, ymax=200,
          legend style={at={(0.95,0.95)}, anchor=north east, legend columns=2},
          ]
          \envelope{brown}{E}{\tiny $P(x_E)$}{supersift.tsv}
          \envelope{orange}{I}{\tiny $P(x_I)$}{supersift.tsv}
        \end{axis}
      \end{tikzpicture}
    }
    \caption{Gatherings, $E$ \& $I$}
  \end{subfigure}
  \caption{
  Superspreading events. Top row, no gatherings.
  This is a relatively slow contagion with $R_0 \approx 1.1$
  Bottom row, periodic gatherings as illustrated in Figure~\ref{fig:gathering}.
  The gatherings result in a much more forceful epidemic with twice the peak infectious population and nearly twice the total infections.
  }
  \label{fig:superspreading}
\end{figure}
The effect of the modest increase in the contact rate is amplified by the partitioning of the population.
Individuals who are gathering interact at this elevated rate only with others who are also gathering.
Being a small fraction of the total population, once one social individual is infected, if they are gathering, the chance of encountering them is proportionally higher: $(0.025N)^{-1} = 40N^{-1}$.
This phenomenon is readily apparent from Figure~\ref{fig:superspreading} where gatherings with the dynamics as described result in a doubling of the peak infectious individuals and a near doubling of the total infections.

\section{Discussion}

This study gives a primer for applying rule-based methods used in molecular biology~\cite{danos_rule-based_2007,chylek_rule-based_2014,kohler_rule-based_2014,bustos_rule-based_2018} to a selection of problems in epidemiological modelling. Each of the models we have chosen would be very challenging to implement as compartmental models because of the scaling issues we discussed. 
They could all be implemented using agent- or individual-based techniques, but, we argue in this paper, not as clearly and parsimoniously as we have done here. 

In fact, the scenarios described above highlight the features of the proposed modelling framework in terms of transparency and compositionality. The scalability of the approach become clear when the examples presented are expressed as reactions.
Table~\ref{tab:scaling} shows, for each example, the number of species or compartments and the number of reactions as well as the number of agents and rules required to capture the same dynamics.
\begin{table}[ht]
  \centering
  \begin{tabularx}{0.6\textwidth}{@{} l | r r | r r @{}}
    model & agents & rules & species & reactions\\
    masks & 1 & 10 & 9 & 96\\
    fomites & 2 & 6 & 11 & 32\\
    vectors & 2 & 6 & 7 & 16\\
    testing & 2 & 10 & 12 & 38\\
    tracing & 3 & 21 & 16 & 170\\
    schools & 1 & 9 & 219 & 89098\\
    gatherings & 1 & 16 & 85 & 766
  \end{tabularx}
  \caption{Number of modelling entities required for the example models given here when treated as rule-based vs reaction-based models. Note that the schools model requires a number of species or compartments and reactions that is in principle unbounded.}
  \label{tab:scaling}
\end{table}

All of the examples are substantially more succinct when written with rules.
The simplest models, of fomites, vector-borne diseases and testing alone are only simpler by a factor of 2 or 3 and could feasibly be studied in reaction-based form.
The other models, despite not being substantially more complex in rule-based form require orders of magnitude more compartments and reactions to capture the same dynamics.

Moreover, the examples provided show that rule-based modelling allows principled expression of interactions in readily-simulated way that is much easier to specify, and allow for greater flexibility in structure.

Rule-based models are also \emph{compositional} meaning that they can be easily combined: with some semantic assumptions, combining models can simply mean concatenating their rules.

Rule-based modelling has previously been applied to address the limitations of traditional approaches for modelling chemical kinetics in cell signalling systems~\cite{danos_rule-based_2007,chylek_rule-based_2014,kohler_rule-based_2014}. 
An attempt to develop a rule-based model for chronic-disease epidemiology has also been made previously~\cite{chiem_rule-based_2012}, but the core methodology there was somewhat intertwined with agent-based models. 
As we have mentioned in the introduction, the approach that we have demonstrated here of constructing and simulating a chemical master equation, is different from agent- or individual-based modelling.
It is also different from reaction-based models that have been considered for epidemiology~\cite{lorton_compartmental_2019} in that it manages combinatorial explosion well. 
Rule-based modelling provides a flexible and computationally efficient methodology that can easily be adapted, and expanded to answer existing and emerging questions in epidemiology. 

The novelty of our work is in translating an established molecular biology modelling framework to epidemiological modelling, with a view to timely application to the COVID-19 epidemic.

The spread of the SARS-CoV-19 virus during 2020 causing a pandemic of COVID-19 across the world, has highlighted the importance of modelling in decision making.
Modelling has been at the forefront of the discussion around imposition of social distancing measures and evaluation of different scenarios to relax them~\cite{ferguson_impact_2020,prem_effect_2020,stutt_modelling_2020,panovska-griffiths_determining_2020,colbourn_modelling_2020,milne_modelling_2020}. 
Having an appropriate model for the available data at every step of the growing epidemic is important and this requires variety of modelling approaches, each with different strengths and weaknesses~\cite{panovska-griffiths_can_2020,jewell_predictive_2020,adam_special_2020}. 
Our contribution with this paper is to highlight another approach in modelling infectious disease spread -- in this case borrowed from molecular biology.   

We note that although we have demonstrated the expressive power of rule-based modelling, the examples given here are simply that: examples and represent a proof-of-principle.
They are intended to show some scenarios that are detailed enough to be interesting but they are simple and consider specific phenomena in isolation.
Each of these examples could usefully be elaborated and studied in greater detail.
Because the mechanisms underlying infectious disease propagation and interventions can be explicitly represented, studying these and other examples as rule-based models is likely to yield important insights.
Rule-based modelling is a powerful tool to gain a more detailed understanding of the dynamics of outbreaks and the options available for their management.
Immediate future work is in applying these techniques to pressing questions from the \textsc{covid-19} epidemic: how the detail of different testing and tracing strategies affects success, the role played by superspreading events, and the interplay between social dynamics and epidemic dynamics.

Rule-based modelling is not a panacea.
There are several practical challenges to its adoption, and some kinds of model that are difficult to express.

First, the notation and approach is not well understood in epidemiology, and this requires a change in practice.
There is potential for misunderstanding where key terminology -- in particular the words ``agent'' and ``compartment'' are used in different senses by the different communities.
We argue that the simplicity and elegance of representation, thinking simply in terms of simple individual interactions rather than the set of complicated differential equations that can be derived for them is sufficient to warrant the use of the rule-based representation.
The benefit of \emph{understandable} models, where a single description is suitable both for computer simulation and human digestion is substantial; it turns opaque science and makes it transparent.
While there is some inertia and there is some cost to adopting this representation, we think the benefits are immense.

Second, this method is not a ``drop-in'' replacement for differential equation models.
For maximum utility, further work on making the use of rule-based models in other systems easier would be valuable.
Most epidemiology packages in Python or R implement a limited set of models, even those that are intended to allow use of varied structures. 
Adopting rule-based modelling is an easy way to make such packages far more extensible. 
It is possible to control stochastic simulations of rule-based models today Python, and it is possible to generate differential equations for solving using GNU Octave, Matlab, Mathematica and Maple.
It would be useful to generate differential equations for solving in Python from rule-based models, and currently we are not aware of any interface for the R language, commonly used in epidemiological modelling.
These are minor practical limitations, easily remedied with some straightforward work and not limitations in principle.

Finally, there are kinds of models that one would like to express that would require extension of existing rule-based modelling tools.
True network models are also difficult to implement as binding sites can only have zero or one bonds.
This means that, considering these bonds as edges in a graph, it is only possible to have vertices of finite degree and it is cumbersome to have vertices of more than very small degree.
Addressing this limitation would require an extension to the core language and cannot be solved with code generation.
Partitions (which we would like to call compartments but that would collide with the use of the word in epidemiology) between which agents are permitted to migrate and within which rules are scoped are cumbersome to express.
The ``gatherings'' model of Section~\ref{sec:gatherings} does this, partitioning the population into those that have gathered together and those that have not, but it is easy to see that this would quickly become unwieldy for a large number of partitions.
Spatial extensions to the $\kappa$ language~\cite{sorokina_simulator_2013} exist that automate the process of generating rules for partitions of this kind.
If the computational expense is tolerable it is possible to conduct rule-based simulations of spatial models.

We have shown that rule-based modelling has a major advantage in expressivity and compositionality over the current practice with compartmental models in epidemiology, and in clarity over individual-based models.
This work brings a broad range of phenomena that are both interesting and important to understand within the scope of what can be studied in a systematic and principled way.
We have demonstrated this by example, providing seven easy pieces: simple, yet interesting models that provide not only an illustration of the utility of rule-based modelling, but starting points for further study.

\paragraph{Acknowledgements}
The authors would like to thank C. Talalaev for the suggestion to model
doorknobs which developed into the hand washing and fomites example.
WW acknowledges support from the Chief Scientist Office grant number
COV/EDI/20/12 and the Medical Research Council (MRC) grant number MR/V027956/1.
JPG acknowledges support from the National Institute for Health Research (NIHR)
Applied Health Research and Care North Thames at Bart’s Health NHS Trust (NIHR
ARC North Thames).

\small
\singlespace
\bibliographystyle{elsarticle-num}
\bibliography{epi}
\normalsize

\appendix
\clearpage
\section{Code: Masks}
\label{code:masks}
\begin{lstlisting}
%var: beta  0.034  // infection probability per contact
%var: c     13     // contact rate per day
%var: alpha 0.2    // progression from exposed to infectious
%var: gamma 0.1429 // progression from infectious to removed
%var: mu    1.0    // convincing rate for mask wearing
%var: nu    0.2    // convincing rate for mask removal
%var: sp    0.5    // spontaneous use of masks

// effect of different combinations of mask wearing
%var: mask_nn 1.0  // no masks
%var: mask_yn 0.5  // infectious wears mask, susceptible not 
%var: mask_ny 0.8  // susceptible wears mask, infectious not
%var: mask_yy 0.2  // both wear masks

%agent: P(x{s e i r} m{y n})

'progression'  P(x{e/i}) @ alpha
'removal'      P(x{i/r}) @ gamma

'infection_nn' P(x{s/e}, m{n}), P(x{i}, m{n}) @ mask_nn * beta * c / N
'infection_yn' P(x{s/e}, m{y}), P(x{i}, m{n}) @ mask_yn * beta * c / N
'infection_ny' P(x{s/e}, m{n}), P(x{i}, m{y}) @ mask_ny * beta * c / N
'infection_yy' P(x{s/e}, m{y}), P(x{i}, m{y}) @ mask_yy * beta * c / N

'convincing'   P(m{n/y}), P(m{y}) @ mu/N // start wearing masks
'removing'     P(m{y/n}), P(m{n}) @ nu/N // stop wearing masks

'spont_mask'   P(m{n/y}) @ sp*PI
'spont_unmask' P(m{y/n}) @ sp*(1-PI)

%obs: S     |P(x{s})|
%obs: E     |P(x{e})|
%obs: I     |P(x{i})|
%obs: R     |P(x{r})|
%obs: M     |P(m{y})|

// variables for population size and initialisation
%var: N       10000       // total population
// initially infectious
%var: INIT_I  50
%var: INIT_S  N - INIT_I
// fraction susceptible
%var: PI      INIT_I/N

%init: INIT_I P(x{i}, m{n})
%init: INIT_S P(x{s}, m{n})

%mod: [true] do $UPDATE PI |P(x{i})|/N; repeat [true]
\end{lstlisting}
\clearpage
\section{Code: Hand washing}
\label{code:fomites}
\begin{lstlisting}
%var: beta  0.025  // infection probability per contact
%var: alpha 0.2    // progression from exposed to infectious
%var: gamma 0.1429 // progression from infectious to removed
%var: omega 8      // wash hands omega times per day
%var: tau   16     // rate at which hands become dirty
%var: kappa 12     // contact rate with surfaces
%var: delta 6      // decontamination rate for surfaces

// individuals can have clean or dirty hands
%agent: P(x{s e i r} h{c d})
// surfaces can be contaminated or not
%agent: S(c{y n})

'progression'     P(x{e/i}) @ alpha
'removal'         P(x{i/r}) @ gamma

'washing'         P(h{d}) <-> P(h{c}) @ omega, tau

'contamination'   S(c{#}), P(x{i}, h{d}) -> S(c{y}), P(x{i}, h{d}) @ kappa/N
'decontamination' S(c{y/n}) @ delta

'infection'       P(x{s/e}, h{d}), S(c{y}) @ beta * kappa / NS

%obs: S     |P(x{s})|
%obs: E     |P(x{e})|
%obs: I     |P(x{i})|
%obs: R     |P(x{r})|
%obs: C     |S(c{y})|

// variables for population size and initialisation
%var: N      10000       // total population
%var: NS     2500        // number of surfaces
// initially infectious
%var: INIT_I 100
// initially susceptible
%var: INIT_S N - INIT_I

%init: INIT_I P(x{i}, h{d})
%init: INIT_S P(x{s}, h{d})
%init: NS     S(c{n})
\end{lstlisting}
\clearpage
\section{Code: Vectors}
\label{code:vectors}
\begin{lstlisting}
%var: beta   0.036   // probability of infection from a bite
%var: bprime 1       // probability of vector becoming infectious
%var: alpha  0.2     // progression from exposed to infectious
%var: gamma  0.1429  // progression from infectious to removed
%var: kappa  1.0     // bites per day per mosquito
%var: kb     0.1429  // birth rate slightly slower than
%var: kd     0.1429  // death rate
%var: water  1       // amount of the breeding habitat available

// individuals simply have disease progression states
%agent: P(x{s e i r})
// vectors can be susceptible or infectious
%agent: V(x{s i})

'progression'     P(x{e/i}) @ alpha
'removal'         P(x{i/r}) @ gamma

'birth'           V(), . -> V(), V(x{s}) @ water * kb
'death'           V()    -> .            @ kd

'biting'          V(x{s/i}), P(x{i}) @ bprime * kappa / M
'infection'       P(x{s/e}), V(x{i}) @ beta * kappa / N

%obs: S     |P(x{s})|
%obs: E     |P(x{e})|
%obs: I     |P(x{i})|
%obs: R     |P(x{r})|
%obs: Vs    |V(x{s})|
%obs: Vi    |V(x{i})|

// variables for population size and initialisation
%var: N      10000       // total population
%var: M      50000       // number of mosquitoes
// initially infectious
%var: INIT_I 100
// initially susceptible
%var: INIT_S N - INIT_I

%init: INIT_I P(x{i})
%init: INIT_S P(x{s})
%init: M      V(x{s})

// perturbation: track the number of vectors
%mod: [true] do $UPDATE M |V()|;            repeat [true]
\end{lstlisting}
\begin{lstlisting}
// perturbation: destroy habitat
%mod: [T] > 60 do $UPDATE water water * 0.5;
\end{lstlisting}
\clearpage
\section{Code: Testing}
\label{code:testing}
\begin{lstlisting}
%var: beta   0.034   // probability of infection from contact
%var: c      13      // contact rate, lower than normal
%var: alpha  0.2     // progression from exposed to infectious
%var: gamma  0.1429  // progression from infectious to removed
%var: theta0 0.0714  // rate of testing in the general population
%var: thetaI 1.0     // rate of testing of infectious population
%var: theta  theta0 + thetaI - theta0*thetaI
%var: m      500     // rate of manufacturing tests
%var: r      0.8     // recall - true positives per positive
%var: s      0.8     // specificity - true negatives per negative

// individuals have disease progression, testability and isolation status
%agent: P(x{s e i r} t{y n} q{y n})
// tests are discrete entities
%agent: T()

'manufacturing' . -> T()  @ m

'progression'   P(x{e/i}, t{n/y}) @ alpha
'removal'       P(x{i/r}, t{y/n}) @ gamma

'infection'     P(x{s/e}, q{n}), P(x{i}, q{n}) @ beta * c / N

'test_tn'       P(t{n}, q{n}), T() -> P(t{n}, q{n}), . @ s*theta0
'test_fp'       P(t{n}, q{n}), T() -> P(t{n}, q{y}), . @ (1-s)*theta0
'test_tp'       P(t{y}, q{n}), T() -> P(t{y}, q{y}), . @ r*theta
'test_fn'       P(t{y}, q{n}), T() -> P(t{y}, q{n}), . @ (1-r)*theta

'exit_s'        P(x{s}, q{y/n}) @ gamma
'exit_r'        P(x{r}, q{y/n}) @ gamma

%obs: Sn    |P(x{s}, q{n})|
%obs: En    |P(x{e}, q{n})|
%obs: In    |P(x{i}, q{n})|
%obs: Rn    |P(x{r}, q{n})|
%obs: Sy    |P(x{s}, q{y})|
%obs: Ey    |P(x{e}, q{y})|
%obs: Iy    |P(x{i}, q{y})|
%obs: Ry    |P(x{r}, q{y})|

// variables for population size and initialisation
%var: N      10000       // total population
// initially infectious
%var: INIT_I 100
// initially susceptible
%var: INIT_S N - INIT_I

%init: INIT_I P(x{i}, t{n}, q{n})
%init: INIT_S P(x{s}, t{n}, q{n})
\end{lstlisting}
\clearpage
\section{Code: Tracing}
\label{code:tracing}
\begin{lstlisting}
%var: beta   0.034   // probability of infection from contact
%var: c      13      // contact rate, lower than normal
%var: alpha  0.2     // progression from exposed to infectious
%var: gamma  0.1429  // progression from infectious to removed
%var: theta0 0.0714  // rate of testing in the general population
%var: thetaI 1.0     // rate of testing of infectious population
%var: theta  theta0 + thetaI - theta0*thetaI
%var: m      100     // rate of manufacturing tests
%var: r      0.8     // recall - true positives per positive
%var: s      0.8     // specificity - true negatives per negative
%var: eta    0.45    // trace 90% of contacts in two days

// individuals have disease progression, testability and isolation status
%agent: P(x{s e i r} t{y n} q{y n})
// tests are discrete entities
%agent: T()
// traces follow the same disease progression as individuals
%agent: C(x{s e i r})

// manufacturing of tests
'manufacturing' . -> T()  @ m

// progression of individuals
'progression'   P(x{e/i}, t{n/y}) @ alpha
'removal'       P(x{i/r}, t{y/n}) @ gamma

// progression of traces
'c_progression' C(x{e/i}) @ alpha
'c_removal'     C(x{i/r}) @ gamma
'c_degradation' C() -> .  @ gamma

// contact rules
'lucky'         P(x{s}, q{n}), P(x{i}, q{n}), . -> 
                P(x{s}, q{n}), P(x{i}, q{n}), C(x{s})   @ (1 - beta) * c / N  
'infection'     P(x{s}, q{n}), P(x{i}, q{n}), . -> 
                P(x{e}, q{n}), P(x{i}, q{n}), C(x{e})   @ beta * c / N
'exposed'       P(x{e}, q{n}), P(x{i}, q{n}), . -> 
                P(x{e}, q{n}), P(x{i}, q{n}), C(x{e})   @ c / N  
'infected'      P(x{i}, q{n}), P(x{i}, q{n}), . -> 
                P(x{i}, q{n}), P(x{i}, q{n}), C(x{i})   @ c / N  
'immune'        P(x{r}, q{n}), P(x{i}, q{n}), . -> 
                P(x{r}, q{n}), P(x{i}, q{n}), C(x{r})   @ c / N  

// testing rules
'test_tn'       P(t{n}, q{n}), T() -> P(t{n}, q{n}), . @ s*theta0
'test_fp'       P(t{n}, q{n}), T() -> P(t{n}, q{y}), . @ (1-s)*theta0
'test_tp'       P(t{y}, q{n}), T() -> P(t{y}, q{y}), . @ r*theta
'test_fn'       P(t{y}, q{n}), T() -> P(t{y}, q{n}), . @ (1-r)*theta

// tracing rules
'trace_s'       P(x{s}, q{#}), P(q{y}), C(x{s}) ->
                P(x{s}, q{y}), P(q{y}), .        @ eta * theta / N
'trace_e'       P(x{e}, q{#}), P(q{y}), C(x{e}) ->
                P(x{e}, q{y}), P(q{y}), .        @ eta * theta / N
'trace_i'       P(x{i}, q{#}), P(q{y}), C(x{i}) ->
                P(x{i}, q{y}), P(q{y}), .        @ eta * theta / N
'trace_r'       P(x{r}, q{#}), P(q{y}), C(x{r}) ->
                P(x{r}, q{y}), P(q{y}), .        @ eta * theta / N

// exit from isolation
'exit_s'        P(x{s}, q{y/n}) @ gamma
'exit_r'        P(x{r}, q{y/n}) @ gamma

%obs: Sn    |P(x{s}, q{n})|
%obs: En    |P(x{e}, q{n})|
%obs: In    |P(x{i}, q{n})|
%obs: Rn    |P(x{r}, q{n})|
%obs: Sy    |P(x{s}, q{y})|
%obs: Ey    |P(x{e}, q{y})|
%obs: Iy    |P(x{i}, q{y})|
%obs: Ry    |P(x{r}, q{y})|
%obs: Cs    |C(x{s})|
%obs: Ce    |C(x{e})|
%obs: Ci    |C(x{i})|
%obs: Cr    |C(x{r})|

// variables for population size and initialisation
%var: N      10000       // total population
// initially infectious
%var: INIT_I 100
// initially susceptible
%var: INIT_S N - INIT_I

%init: INIT_I P(x{i}, t{n}, q{n})
%init: INIT_S P(x{s}, t{n}, q{n})
\end{lstlisting}
\clearpage
\section{Code: Schools}
\label{code:school}
\begin{lstlisting}
%var: beta   0.034  // infection probability per contact
%var: c      13     // contacts per unit time
%var: d      0.01   // contact duration ~ 15 mins
%var: alpha  0.2    // progression from exposed to infectious
%var: gamma  0.1429 // progression from infectious to removed
%var: school 0.5    // fraction of the waking day spent at school
%var: lock   0.3    // reduction in contact due to lockdown
%var: child  2      // increase in contact at school

// individuals have the usual disease progression states and
// are in two groups -- children and adults. bonds form between
// these individuals.
%agent: P(x{s e i r} a{c a} e1 e2 e3)

// Pair adults
'parents'  P(a{a}, e3[./3]), P(a{a}, e3[./3]) @ inf

// Produce children
'children' P(a{a}, e1[./1], e3[3]), P(a{a}, e2[./2], e3[3]),
           P(a{c}, e1[./1], e2[./2]) @ inf

// infection 
'infection'    P(x{s/e}), P(x{i})               @ lock*beta*c/N
'infection_sc' P(x{s/e}, a{c}),  P(x{i}, a{c})  @ (child*school*beta*c)/(P_CHILD*N)
'infection_e1' P(x{s/e}, e1[1]), P(x{i}, e1[1]) @ (1-school)*beta/d
'infection_e2' P(x{s/e}, e2[2]), P(x{i}, e2[2]) @ (1-school)*beta/d
'infection_e3' P(x{s/e}, e3[3]), P(x{i}, e3[3]) @ beta/d

'progression' P(x{e/i}) @ alpha
'removal'     P(x{i/r}) @ gamma

%obs: S     |P(x{s})|
%obs: E     |P(x{e})|
%obs: I     |P(x{i})|
%obs: R     |P(x{r})|
%obs: Sa    |P(a{a}, x{s})|
%obs: Ea    |P(a{a}, x{e})|
%obs: Ia    |P(a{a}, x{i})|
%obs: Ra    |P(a{a}, x{r})|
%obs: Sc    |P(a{c}, x{s})|
%obs: Ec    |P(a{c}, x{e})|
%obs: Ic    |P(a{c}, x{i})|
%obs: Rc    |P(a{c}, x{r})|
%obs: F1    |P(e3[.])|
%obs: F2    |P(e1[.],e2[.],e3[_])|/2
%obs: F3    |P(e1[_],e2[.],e3[_])|
%obs: F4    |P(e1[_],e2[_],e3[_])|/2

// variables for population size and initialisation
%var: N      10000       // total population
// initially infectious
%var: INIT_I 50
// initially susceptible
%var: INIT_S N - INIT_I

// probability of being a child
%var: P_CHILD 0.25

%init: INIT_I*P_CHILD     P(x{i}, a{c})
%init: INIT_I*(1-P_CHILD) P(x{i}, a{a})
%init: INIT_S*P_CHILD     P(x{s}, a{c})
%init: INIT_S*(1-P_CHILD) P(x{s}, a{a})
\end{lstlisting}
\clearpage
\section{Code: Gatherings}
\label{code:super}
\begin{lstlisting}
%var: beta     0.016  // infection probability per contact
%var: c_n      10     // normal contact rate per day
%var: c_g      100    // gathering contact rate per day
%var: alpha    0.2    // progression from exposed to infectious
%var: gamma    0.1429 // progression from infectious to removed
%var: gather   0      // how often gathering happens
%var: k_gather 4      // how frequently to gather
%var: k_leave  24     // how long a gathering lasts

%agent: P(x{s e i r} g{y n} c{=1 / +=10})

'progression'  P(x{e/i}) @ alpha
'removal'      P(x{i/r}) @ gamma

'infection_n'  P(x{s/e}, g{n}), P(x{i}, g{n}) @ beta * c_n / NN
'infection_g'  P(x{s/e}, g{y}), P(x{i}, g{y}) @ beta * c_g / NG

'gathering'    P(g{n/y}, c{=social}) @ gather * k_gather * social/10
'leaving'      P(g{y/n}) @ k_leave

'sort_1'  P(c{=0/+=1})  @ inf
'sort_2'  P(c{=0/+=2})  @ inf
'sort_3'  P(c{=0/+=3})  @ inf
'sort_4'  P(c{=0/+=4})  @ inf
'sort_5'  P(c{=0/+=5})  @ inf
'sort_6'  P(c{=0/+=6})  @ inf
'sort_7'  P(c{=0/+=7})  @ inf
'sort_8'  P(c{=0/+=8})  @ inf
'sort_9'  P(c{=0/+=9})  @ inf
'sort_10' P(c{=0/+=10}) @ inf

%obs: S     |P(x{s})|
%obs: E     |P(x{e})|
%obs: I     |P(x{i})|
%obs: R     |P(x{r})|
%obs: G     |P(g{y})|
%obs: P1    |P(x{r}, c{=1})|
%obs: P2    |P(x{r}, c{=2})|
%obs: P3    |P(x{r}, c{=3})|
%obs: P4    |P(x{r}, c{=4})|
%obs: P5    |P(x{r}, c{=5})|
%obs: P6    |P(x{r}, c{=6})|
%obs: P7    |P(x{r}, c{=7})|
%obs: P8    |P(x{r}, c{=8})|
%obs: P9    |P(x{r}, c{=9})|
%obs: P10   |P(x{r}, c{=10})|

// variables for population size and initialisation
%var: N       10000       // total population
// initially infectious
%var: INIT_I  50
%var: INIT_S  N - INIT_I

%init: INIT_I P(x{i}, g{n}, c{=0})
%init: INIT_S P(x{s}, g{n}, c{=0})

%var: NN N
%var: NG 0
%mod: [true] do $UPDATE NN |P(g{n})|; repeat [true]
%mod: [true] do $UPDATE NG |P(g{y})|; repeat [true]

%mod: alarm 0.25 gather > 0 do $UPDATE gather 0; repeat [true]
%mod: alarm 1               do $UPDATE gather 1; repeat [true]
\end{lstlisting}

\end{document}